\documentclass[useAMS,usenatbib]{mn2e}

\usepackage{psfig, epsf, epsfig}

\title[Interstellar molecules in galaxies]{Simulating
the spatial distributions of gas- and ice-phase  molecules in galaxies: 
a new method and preliminary results}
\author[K. Bekki, K. Furuya, and T. Shimonishi]
{Kenji Bekki${}^1$\thanks{E-mail:
kenji.bekki@uwa.edu.au},
Kenji Furuya${}^{2,3}$,
and Takashi Shimonishi${}^4$ \\
${}^1$ICRAR M468
The University of Western Australia
35 Stirling Hwy, Crawley
Western Australia, 6009 \\
${}^2$
Department of Astronomy, Graduate School of Science, University of Tokyo, Tokyo 113-0033, Japan 
 \\
${}^3$
The Institute of Physical and Chemical Research (RIKEN), 2-1, Hirosawa, Wako-shi, Saitama 351-0198, Japan \\
${}^4$
Institute of Science and Technology, Niigata University, Ikarashi-ninocho 8050, Nishi-ku, Niigata 950-2181, Japan
}

\begin{document}

\date{Accepted, Received 2005 February 20; in original form }

\pagerange{\pageref{firstpage}--\pageref{lastpage}} \pubyear{2005}

\maketitle

\label{firstpage}

\begin{abstract}

Recent observations have revealed significant variations in the abundances of gas- and ice-phase 
molecules in galaxies with different luminosities
and types. 
In order to discuss the physical origins of these variations, we
incorporate gas- and dust-phase interstellar chemistry into  galaxy-scale simulations with 
various baryonic physics including dust formation, evolution, and destruction, all of which
are essential for the calculations of $\approx 400$ interstellar molecule species.
The new simulations can accordingly predict the abundances of gas- and ice-phase  molecular species 
such as ${\rm H_2O}$ and ${\rm CO_2}$ ice
within individual molecular gas cloud  of galaxies 
based on gas density and temperature, dust temperature ($T_{\rm dust}$),
elemental abundances (e.g., CHNOPS),
UV radiation strength ($F_{\rm UV}$),  and cosmic ray ionisation rate ($\zeta_{\rm CR}$) within the clouds.
Since this is the first of the series of papers,
we describe the details of the new simulations and present the preliminary results focused on
the spatial distributions of ${\rm H_2O}$, CO, ${\rm CO_2}$, and ${\rm CH_3OH}$ ice species
in a disk galaxy similar to the Milky Way.
We particularly discuss how 
$T_{\rm dust}$ and gas-phase elemental abundances
can control the spatial distributions of the above molecules in galaxies.
We briefly discuss the total amount of ${\rm H_2O}$ and ${\rm CO_2}$ ices and radial distributions
of PN and PO molecules in  the Galaxy.
\end{abstract}

\begin{keywords}
Galaxies -- astrochemistry -- dust, extinction -- galaxies: abundances -- galaxies: star formation
\end{keywords}

\section{Introduction}

Interstellar dust plays various roles in the formation of galaxies, star clusters,  and planets
on multiple physical scales,
such as formation of low-mass stars in the early universe (e.g., Schneider et al. 2012;
Chiaki et al. 2015),
thermal balance in interstellar medium (ISM) and collapsing
gas clouds (e.g., McKee 2011; Sharda \& Krumholz 2022),
severe suppression of galaxy-wide star formation due to photoelectric heating of dust
(e.g., Forbes et al. 2015; Osman et al. 2020),
co-evolution of dust and metals in galaxies (e.g., Asano et al. 2013; Bekki 2013, B13),
${\rm H_2}$ formation on dust grains (e.g.,  Gould \& Salpeter 1963; Hirashita 2023),
formation of various molecules on the surface of dust 
(e.g., Greenberg et al. 1972; Watson \& Salpeter 1972; Hama \& Watanabe 2013),
and formation of planets from aggregation of tiny dust particles (e.g, Hayashi et al. 1985).
Among these roles, interstellar chemistry on dust grains is fundamentally important
in discussing the possibly diverse origins of various molecules observed in different phases of
ISM within galaxies 
(e.g., Herbst \& van Dishoeck 2009).
Thus, theoretical studies of galaxy formation and evolution need to incorporate
both dust formation and evolution and interstellar chemistry in their models self-consistently
in order to discuss
the physical origins of the observed rich variety of molecules in galaxies.

In the last decade, there has been great progress in observations of various molecular species in 
regions within galaxies with different metallicities 
(see Shimonishi et al. 2025 for a review).
For example, single-dish radio observations have revealed giant-molecular-cloud-scale ($>$10 pc) chemical compositions 
in external low-metallicity galaxies such as the Large and Small Magellanic Clouds (LMC and SMC, respectively)
(Johansson et al. 1994; Chin et al. 1997; Heikkil et al. 1999; Wang et al. 2009;
Paron et al. 2016; Nishimura et al. 2016a; Galametz et al. 2020; Gong et al. 2023),
IC10 (Nishimura et al. 2016b; Kepley 2018),
and NGC6822 (Braine et al. 2017).
In addition, radio observations of molecular-cloud-scale ($>$several pc) chemistry are reported for the 
outer part of our Galaxy 
(Ruffle et al. 2007; Braine et al. 2007; Blair et al. 2008; Bernal et al. 2021; Fontani et al. 2022a;
Braine et al. 2023).
Metallicities of these regions are lower than the solar value by a factor of about two to ten, suggesting that the chemical evolution assisted by dust grains would be different compared to that in the solar neighborhood.
In fact, for example, enhanced photochemistry is suggested in low-metallicity molecular clouds based on the increase in the abundance of a radical species, CCH, which would be the consequence of the reduced shielding of the interstellar radiation field (ISRF) 
owing to the low dust abundance (e.g., Nishimura et al. 2016a).

High-spatial-resolution radio observations with the Atacama Large Millimetre/submillimetre Array (ALMA) have revealed molecular line emission from dense ($n_{\mathrm{H_2}}$ $>$10$^6$ cm$^3$), compact ($<$0.1 pc), and high-temperature (T $>$100 K) gas associated with protostars (known as hot cores or hot corinos) located in distant 
low-metallicity regions, such as the LMC 
(e.g., Shimonishi et al. 2016b; Shimonishi et al. 2020; Sewi{\l}o et al. 2018, 2022a, 
Hamedani Golshan et al. 2024),
the SMC (Shimonishi et al. 2023), and the outer Galaxy (Shimonishi et al. 2021).
A variety of complex organic molecules (COMs) 
have been detected in protostellar cores in dust-poor regions.
The detection of hot cores (including candidates) in the Galactic centre 
(a high-metallicity environment) 
is reported by Bonfand et al. (2017), Miyawaki et al. (2021), and Jeff et al. (2024).
The origin of these COMs in various galaxy environments is still under debate.

Infrared observations have revealed chemical compositions of 
ices associated with deeply embedded protostars in the LMC
(e.g., van Loon et al. 2005; Shimonishi et al. 2008, 2010, 2016; Oliveira et al. 2009; Seale et al. 2011),
and the SMC  (e.g., van Loon et al. 2008;  
Oliveira et al. 2011, 2013, Shimonishi et al. 2018).
For example,
the CO$_2$/H$_2$O ice ratios in LMC sources are higher on average and show a larger scatter compared to Galactic sources.
The CO/H$_2$O ice ratios are reported to be similar between LMC and Galactic sources, while CH$_3$OH/H$_2$O ice ratios are lower in LMC sources.
Furthermore,
ice observations are also reported for the Galactic centre regions with high metallicities
(Chiar et al. 2000, 2002; An et al. 2009, 2011).
The physical origins of these ices in different types of galaxies  would need to be 
addressed by theoretical models of the galaxies.


\begin{table*}
	\centering
	\caption{  A brief summary of previous observations of interstellar molecules in the Galaxy and external galaxies. 
 Molecular abundances of gas and ice have been measured for various objects (embedded protostars, EP; hot molecular cores, HMC; molecular clouds; MC; giant molecular clouds, GMC) in diverse galactic environments (Galactic center, GC; outer Galaxy, OG; Large Magellanic Cloud, LMC; Small Magellanic Cloud, SMC; and other external galaxies). 
    Note that molecular observations for nearby regions are not included in the table. 
    }
	\label{tab:example_table}
	\begin{tabular}{lcccr} 
		\hline
{ Molecular species}
& { Phase}
& { Observed objects }
& { Regions } 
& { Literature example} \\
        \hline
CO, ${\rm HCO^{+}}$, HCN, CS, CCH, COMs etc &
Gas & 
MC & 
GC & 
Armijos-Abenda\~no et al. (2015) \\
${\rm H_2O}$, ${\rm CO_2}$, ${\rm CH_3OH}$ &
Ice & 
EP & 
GC & 
An et al. (2011, 2017) \\
%
CO, CS, SO etc  &
Gas & 
MC & 
OG &
Ruffle et al. (2007) \\
${\rm H_2CO}$  &
Gas & 
MC & 
OG &
Blair et al. (2008) \\
${\rm HCO^{+}}$, HCN, ${\rm CH_3OH}$ etc &
Gas & 
MC & 
OG &
Fontani et al. (2022ab) \\
%
%
HDO, ${\rm CH_3OH}$, ${\rm CH_3CN}$, ${\rm SO_2}$, COMs etc. &
Gas & 
HMC & 
OG &
Shimonishi et al. (2021) \\
CO, ${\rm HCO^{+}}$, HCN, CS, SO etc &
Gas & 
MC &
LMC &
Nishimura et al. (2016a) \\
%
${\rm H_2O}$, CO, ${\rm CO_2}$, ${\rm CH_3OH}$ &
Ice & 
EP & 
LMC & 
Shimonishi et al. (2016a) \\
${\rm CH_3OH}$, ${\rm CH_3CN}$, ${\rm SO_2}$, CCH, CN etc &
Gas & 
HMC & 
LMC &
Shimonishi et al. (2020) \\
${\rm CH_3OH}$, COMs etc &
Gas & 
HMC & 
LMC &
Sewilo et al. (2020) \\
%
${\rm HCO^{+}}$, HCN, CN etc  &
Gas & 
MC &
SMC &
Chi et al. (1998) \\
%
${\rm H_2CO}$, ${\rm CH_3OH}$, ${\rm SO_2}$, ${\rm HCO^{+}}$ etc &
Gas & 
HMC & 
SMC &
Shimonishi et al. (2023) \\
${\rm H_2O}$, ${\rm CO_2}$ &
Ice & 
EP & 
SMC & 
Oliveira et al. (2013) \\
%
${\rm HCO^{+}}$, HCN, HNC, CS, SO, CCH &
Gas & 
GMC  &
IC10 & 
Nishimura et al. (2016b) \\
${\rm HCO^{+}}$, HCN, CS, SO, CCH &
Gas & 
GMC &
NGC6822 & 
Braine et al. (2017) \\
${\rm HCO^{+}}$, HCN, CCH  &
Gas & 
GMC &
M33 & 
Braine et al. (2017) \\
${\rm H_2O}$, ${\rm CO_2}$ &
Ice & 
GMC &
M82 &
Yamagishi et al. (2015) \\
${\rm H_2O}$  &
Gas & 
--  & 
high-z &
Perrotta et al. (2023) \\
${\rm H_2O}$, CO, ${\rm CO_2}$ &
Ice & 
--  & 
high-z &
Sajina et al. (2009, 2025) \\
        \hline
    \end{tabular}
    \begin{flushleft}
    Note. This is a list of selected references only. A more comprehensive list of relevant papers is given in Shimonishi 2025.
    \end{flushleft}
\end{table*}

These observed diverse properties of interstellar molecules 
can possibly reflect the diverse
physical conditions of local ISM
in galaxies, such as dust temperatures
($T_{\rm dust}$), gas densities ($\rho_{\rm g}$),
strengths of interstellar radiation fields (ISFR) and cosmic rays (CR),  and 
gas- and dust-phase abundances of
various elements such as CNO.
These ISM conditions can be controlled by various galaxy-scale dynamics and hydrodynamics,
such as flattening of radial metallicity gradients by dynamical actions of stellar bars
(e.g., Friedli et al. 1995; Fraser-McKelvie et al. 2019),
enhanced or suppressed star formation by galaxy interaction and merging 
(e.g., Noguchi \& Ishibashi 1986; George  2019),
concentration and merging of cold gas clouds driven by dynamical
action of spiral arms (e.g., Koda et al. 2009; Dobbs et al. 2015),
and collisions of molecular clouds and the subsequent formation of massive stars and star clusters
(e.g.,  Bekki et al. 2004; Tsuge et al. 2019).

Thus, theoretical predictions of interstellar molecule abundances 
at different locations in galaxies 
need to be based on 3D numerical simulations that include various dynamical
and hydrodynamical processes of galaxies in a self-consistent manner and accordingly
predict spatial and temporal
variations of their ISM.
However previous galaxy-scale numerical simulations have been unable to
predict the abundances of interstellar molecules in ISM of galaxies with different masses
and types, because interstellar chemistry has not been properly incorporated in the simulations.
The lack of theoretical predictions of interstellar molecular abundances based on 3D dynamical and
hydrodynamical simulations of galaxies has hampered the physical interpretations
of the observational results on the diverse molecular abundances in
galaxies.

It is therefore crucial to 
predict interstellar molecular abundances by applying the astrochemistry
models to the results of galaxy-scale simulations (``post-processing'').
Given that the surface chemistry on interstellar dust is one of key processes in
interstellar chemistry, 
the physical properties of dust such as abundances and size distributions of dust 
and their time evolution
need to be properly predicted by galaxy-scale simulations to discuss the 
observed diverse molecular  properties of galaxies.
Our recent galaxy-scale computer simulations using
smooth particle hydrodynamics (SPH)
have already incorporated 
various dust-related physical processes  in ISM
to predict temporal and spatial variations of gas and  dust properties
 (B13, Bekki 2015a, b;
B15a and B15b, respectively).
It is therefore straightforward and practical
for the present study to 
predict the physical properties of molecules in galaxies through
applications of an astrochemistry code to the results of
galaxy-scale simulations.
Such new simulations with interstellar chemistry
have great advantages in predicting the spatial distributions
of interstellar molecules and thus comparing the prediction with recent
observations mentioned above.

The purpose of this series of papers is to thus investigate (i) the spatial distributions of
various interstellar molecules in galaxies
and (ii) how the molecular abundances are sensitive to the physical conditions of molecular clouds
using new galaxy-scale numerical simulations
that incorporate interstellar chemistry self-consistently.
We will investigate these two points by applying the astrochemistry code by Furuya
et al. (2015) to the 
the simulated physical properties  of star-forming molecular clouds
(e.g., dust temperature, $T_{\rm d}$): This study is based on the so-called
``post-processing'' analysis of simulation data.
The new simulations can predict the abundances of interstellar molecules formed 
from gas-gas and gas-dust chemistry on dust grains in different molecular clouds
at different locations within galaxies with different luminosities and types.
We focus particularly on the abundances of CO gas 
and  ${\rm H_2O}$, ${\rm CO_2}$, and ${\rm CH3OH}$
ice species in the present study:
we will discuss the origin of other species in our forthcoming papers.

The plan of the paper is as follows.
We describe how we predict galaxy-scale distributions of various molecules
by using the interstellar chemistry models
in \S 2.
We present the results of the spatial distributions of 
interstellar molecules in disk galaxies  and describe how the results depend
on the input parameters of the adopted models
in \S 3.
We discuss the physical implications of these new results
in \S 4.
We summarise the main results from the present simulations.
in \S 5.
Table 1 briefly summarises the previous observational results on the physical
properties of various interstellar molecules   which the present 
and our future simulations try to reproduce.

\begin{figure*}
\psfig{file=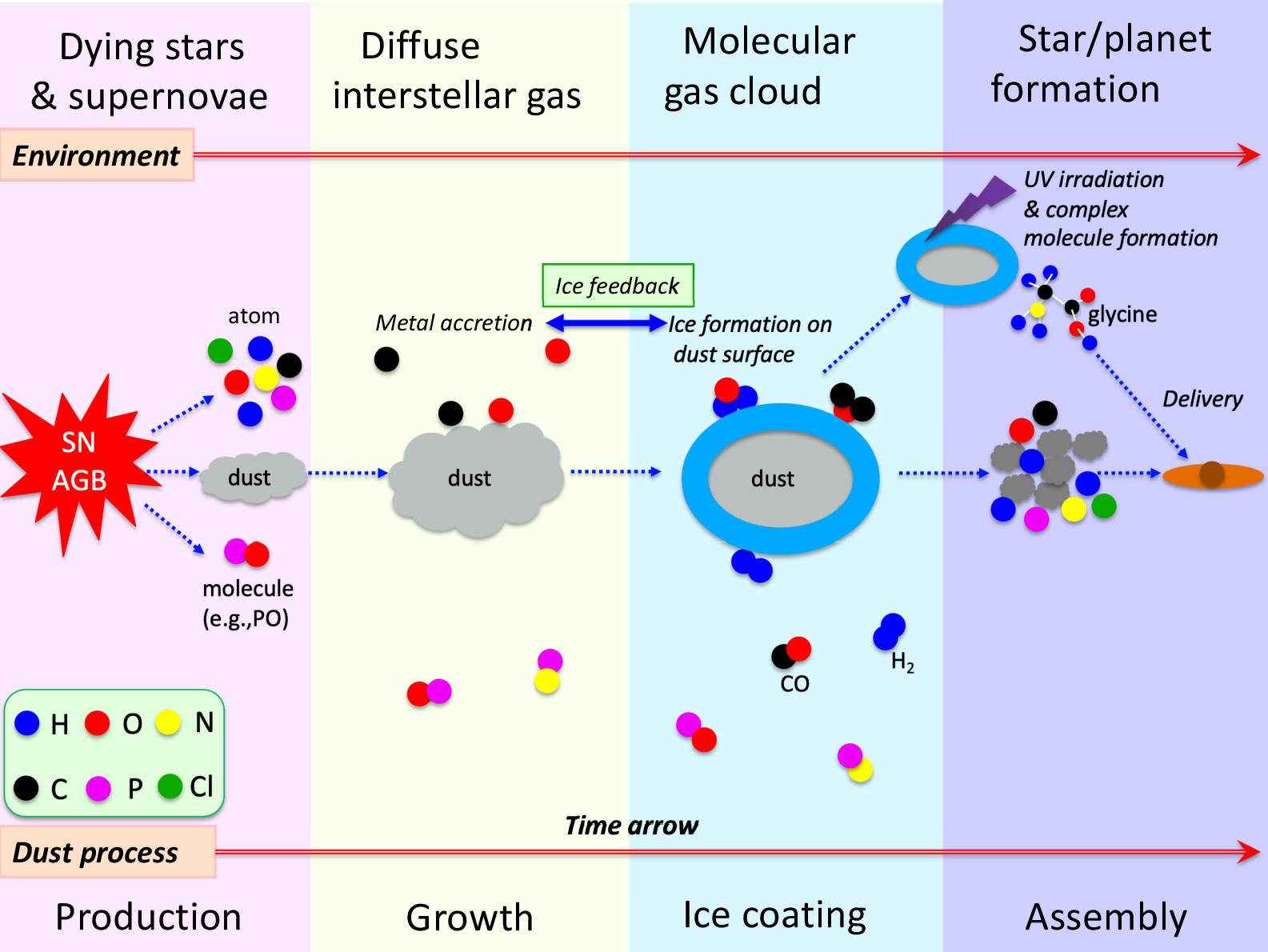,width=13.0cm,angle=90}
\caption{
Illustration of key physical processes that are modelled in the present simulations with a new
code. 
The journey of a dust grain  within a galaxy is 
divided into the following  environments depending on their  physical properties:
(i) dying stars such as AGB stars, core collapse supernovae (CCSNe), and novae
(ii) diffuse interstellar medium, (iii) dense cores of molecular clouds, and (iv) forming
stars and planets. In these environments, various dust formation and evolution processes
can occur, including dust formation through nucleation within gaseous ejecta from CCSNe and AGB stars
in (i),
dust growth through accretion of metals in (ii),  ice formation on the surface of dust grains
in (iii), 
and complex organic matter formation 
due to UV irradiation from a baby star in (iv).
Glycine is shown in (iv) as a possible example of organic matter formation,
though it is yet unclear whether  it is produced in ISM.
Ice formation on dust can possibly suppress the dust growth, whereas dust growth can 
promote ice formation on dust surface. This mutual interaction between ice and dust is called
``ice feedback'' in this illustration. Tiny dust particles merge together
to finally form a planet, and the complex organic matter can be delivered to the planet.
The baby star finally dies away so that its matter can be again circulated to the ISM.
}
\label{Figure. 1}
\end{figure*}

\section{The model}
\subsection{Outline}
Fig. 1 illustrates various key processes of gas, dust, and metals that need to be
incorporated into the present simulations to derive the molecular abundances and their
spatial and temporal variations in the Galaxy and galaxies. It should be stressed here that
the present study does not model a number of physical processes in ISM, such
as dust size evolution,  dust sputtering by hot and shocked gas,
and interstellar chemistry dependent on
dust sizes, and molecular formation in the circumstellar regions of evolved stars
(e.g., red giants and AGB stars).  Although these will be incorporated into our future simulations, 
we only briefly discuss how these processes can possibly influence the present new results
on the molecular abundances in galaxies.

We investigate the spatial distributions of interstellar molecules in 
simulated  disk galaxies based on the distributions of 
gas density and temperature ($\rho_{\rm g}$ and $T_{\rm g}$, respectively),
dust temperature ($T_{\rm dust}$), strengths of local ISRF
($F_{\rm ISRF}$) and  UV radiation  ($F_{\rm UV}$),
cosmic ray ionisation rate ($\zeta_{\rm CR}$), and gas-phase and dust-phase chemical abundances.
Since this is the very first paper in a series of papers,
we focus  particularly  on  the spatial distributions
of molecules in luminous disk galaxies like the Galaxy: We will discuss how the distributions
can be different in galaxies with different masses and types in our forthcoming papers. 
In the present study,
we predict the abundances of various molecular species in simulated galaxies
from the physical properties of ISM at selected time steps only after simulations are completed. 
This ``post-processing'' is the best way for the present study to investigate the abundances
of molecules and their spatial variations, because it is extremely time-consuming
to investigate the molecular abundances for all gas particles ($N \approx 10^6$) at all time steps
during a simulation.

One of disadvantages in this post-processing method is that possibly key interplay
between ice and dust in ISM of galaxies cannot be self-consistently investigated.
For example,
the present simulations do not incorporate the influences of interstellar ice on 
the surfaces of dust grains
on dust growth (e.g., Ferrara et al. 2016; Ceccarelli et al. 2018),  which could be quite important
for the evolution of  dust abundances thus for  the abundances of molecules.
Co-evolution of ice and dust therefore needs to be properly modelled in our future
simulations so that the time evolution of molecular abundances can be more accurately
predicted. Such simulations will be able to be used to predict
both the spatial distributions of various dust grains and those of  ice components
within galaxies  in a fully self-consistent manner.

\subsection{Isolated disk galaxies}

Since isolated disk galaxy models in the present study are the same as those in Bekki (2014),
we only briefly describe them.
A disk  galaxy is assumed to  consist of dark matter, stars, and gas,
and gas-phase and dust-phase metals in interstellar medium (ISM)
 are simply referred to as ``metals'' and ``dust'' 
(i.e., separate components in ISM).
The total masses of dark matter halo, stellar disk, gas disk,
bulge in  a disk galaxy are denoted as $M_{\rm dm}$, $M_{\rm s}$, $M_{\rm g}$,
and $M_{\rm b}$, respectively.
The total masses of  atomic (H{\sc i}) and molecular hydrogen (${\rm H_2}$) 
and dust in a disk galaxy are denoted as
$M_{\rm HI}$, $M_{\rm H_2}$, and $M_{\rm dust}$, respectively.
In order to describe the radial density profile of the dark matter halo in a disk galaxy,
we adopt the density distribution of the NFW
halo (Navarro, Frenk \& White 1996) suggested from CDM simulations:
\begin{equation}
{\rho}(r)=\frac{\rho_{0}}{(r/r_{\rm s})(1+r/r_{\rm s})^2},
\end{equation}
where  $r$, $\rho_{0}$, and $r_{\rm s}$ are
the spherical radius,  the characteristic  density of a dark halo,  and the
scale
length of the halo, respectively.
Using the $c-M_{\rm dm}$ relation
predicted by recent cosmological simulations
(e.g., Neto et al. 2007),
we appropriately choose 
the $c$-parameter ($c=r_{\rm vir}/r_{\rm s}$, where $r_{\rm vir}$ is the virial
radius of a dark matter halo) and $r_{\rm vir}$
for a given $M_{\rm dm}$.

The radial ($R$) and vertical ($Z$) density profiles of a  stellar disk with a disk size $R_{\rm s}$ are
proportional to $\exp (-R/R_{0}) $ with scale
length $R_{0} = 0.2R_{\rm s}$  and to ${\rm sech}^2 (Z/Z_{0})$ with scale
length $Z_{0} = 0.04R_{\rm s}$, respectively.
The gas disk with a size  $R_{\rm g}=2R_{\rm s}$
is also assumed to have a scale length of $R_{0, g}$ and
a vertical scale length $Z_{0,g}$.
The bulge component of a disk galaxy  has a  $R_{\rm b}$
and a scale-length $R_{\rm 0, b}$
and is represented by the Hernquist
density profile.
In addition to the
rotational velocity caused by the gravitational field of stellar and gaseous disk,
bulge, and dark halo components, the initial radial and azimuthal
velocity dispersions are assigned to the disc component according to
the epicyclic theory with Toomre's parameter $Q$ = 1.5.
The vertical velocity dispersion at a given radius is set to be 0.5
times as large as the radial velocity dispersion at that point.

In the present study, we investigate a disk model (``fiducial model'')  with 
$M_{\rm dm}=1.0 \times 10^{12} {\rm M}_{\odot}$, 
$M_{\rm s}=6 \times 10^{10} {\rm M}_{\odot}$, 
$M_{\rm g}=6 \times 10^{9} {\rm M}_{\odot}$, 
$M_{\rm b}=1 \times 10^{10} {\rm M}_{\odot}$, 
$r_{\rm vir}=234$ kpc,
$c=10$,
$R_{\rm s}=17.5$ kpc,
$R_{\rm 0}=R_{\rm 0, g}=3.5$ kpc,
$Z_{\rm 0}=Z_{\rm 0, g}=0.35$ kpc,
$R_{\rm b}=3.5$ kpc,
and $R_{\rm0, b}=0.7$ kpc.
These parameters are chosen such that the structural and kinematical properties
can be very similar to those of our Milky Way (MW) galaxy.
Since the disk galaxy is assumed to have a negative metallicity gradient 
(as described later),
the inner and outer parts of the disk can have quite different metallicities thus
different abundances of interstellar molecules.
We accordingly investigate  the radial profiles of 
the abundances of interstellar molecules
in this MW-type disk galaxy.
The abundances of various interstellar molecules
have been observed in ISM with a wide variety of different physical conditions within
the Galaxy. Therefore  direct comparisons between the observations and the predictions from
the simulations can be made. 
We will investigate the abundances of interstellar molecules in other environments
such as low-metallicity dwarfs, 
in our future studies.
The mass and spatial resolutions for gas particles are $6 \times 10^3 {\rm M}_{\odot}$ and 85 pc,
respectively, in the fiducial model.

\subsection{Star formation and IMF}

We model galaxy-wide star formation based on the
the observed  Kennicutt-Schmidt law, i.e., 
SFR$\propto \rho_{\rm g}^{\alpha_{\rm sf}}$  (Kennicutt 1998),
where $\alpha_{\rm sf}$ is the slope of the power-law.
Here ``star formation'' means 
conversion from gas particles into new collisions stellar particles
(simply referred to as ``new star'') in simulated galaxies.
Star formation can occur
if the following three conditions (i)-(iii) for star formation are satisfied:
(i) the local dynamical time scale is shorter
than the sound crossing time scale (mimicking
the Jeans instability) , (ii) the local velocity
field is identified as being consistent with gravitationally collapsing
(i.e., div {\bf v}$<0$),
and (iii) the local density exceeds a threshold density for
star formation ($\rho_{\rm th}$).
This check of gas-to-star-conversion is done at every $0.01 t_{\rm unit}$
(corresponding to the maximum time step width, $\Delta t_{\rm max}$),
where $t_{\rm unit}$ is the time units ($1.41 \times 10^8$ yr)
adopted in the present simulations. In our previous simulations,
gas-to-star-conversion is done whenever the above three
conditions are satisfied. 
As modelled in our previous works (Bekki \& Shioya 1998; B15a),  
a gas particle is assumed to  have the following probability of
star formation ($P_{\rm sf}$);
\begin{equation}
P_{\rm sf}=1-\exp ( -C_{\rm sf} f_{\rm H_2}
\Delta t_{\rm max} {\rho}^{\alpha_{\rm sf}-1} ),
\end{equation}
where $C_{\rm sf}$ is a normalisation constant,
$f_{\rm H_2}$ is the ${\rm H_2}$ mass fraction of the gas particle,
$\Delta t_{\rm max}$ is the maximum time step width for the particle,
$\rho$ is the gas density of the particle,
and $\alpha_{\rm sf}$ is
the power-law slope of the  Kennicutt-Schmidt law.
The constant $C_{\rm sf}$ 
is chosen such that the mean  galaxy-wide SFR over
1 Gyr evolution of the present
MW model can be $\approx 1 {\rm M}_{\odot}$ as observed:
It can correspond to a star formation efficiency
within cores of molecular clouds.

We adopt the following power-law function for the stellar initial mass function (IMF):
\begin{equation}
\Psi (m_{\rm s}) = C_{\rm imf}{m_{\rm s}}^{-\alpha},
\end{equation}
where $m_{\rm s}$ is the initial mass of
each star and $\alpha$ is the power-law slope of the IMF:$\alpha =2.35$
corresponds to the canonical Salpeter IMF  (Salpeter 1955).
The normalisation factor $C_{\rm imf}$ is determined by  $\alpha$,
$m_{\rm l}$ (lower mass cut-off), and  $m_{\rm u}$ (upper mass cut-off).
These $\alpha$,  $m_{\rm l}$, and $m_{\rm u}$
are  set to be 2.35,   $0.1 {\rm M}_{\odot}$,
and  $50 {\rm M}_{\odot}$, respectively, for all models in the present study.
Although the present results would depend strongly on the adopted IMF parameters,
we discuss the models with the canonical IMF only.
This is  because one of our main purposes
is to describe how galaxy-scale distributions of gas- and ice-phase molecules
can be predicted through applications of our astrochemistry models (F15)
to data from galaxy-scale simulations (i.e., post-processing).

\subsection{Chemical enrichment}
We first  derive the IMF-weighted yields of stars
using the chemical yield tables for CCSNe and SNe Ia provided by 
Tsujimoto et al. (1995; T95) 
and those for AGB stars by
van den Hoek \& Groenewegen (1997,VG97)
for the adopted  standard Salpeter IMF.
We then use the yields to calculate the amount of each metal (e.g., CNO) ejected from 
a stellar particle at different stellar lifetimes 
to chemically pollute the surrounding gas particles: the details of this
enrichment model is given in B13. 
These chemical yields from T95 and VG97 are used to calculate the dust production
rates in CCSNe, SNe Ia, and AGB stars too, as done in B13 and B15a.
We need to adopt the yield of N and those of P and Cl from Bekki \& Tsujimoto (2023)
and Bekki \& Tsujimoto (2024, BT24), respectively,
because B13 did not investigate the chemical evolution of these elements.
We adopt (i) the radial metallicity gradients of initial stellar and gaseous
disks that are consistent  with the observed ones for star clusters
in the Galaxy (B15a) 
and  (ii) the mean age of 5 Gyr for old stars initially in the stellar disk.

We investigate the spatial distributions of P-bearing molecules within galaxies
in the present and future studies.
Therefore  we include the detailed model for P enrichment by a variety of stars
including SNe, AGB stars, and novae  based on
the results in BT24.  
The major production site of P in galactic chemical evolution
is oxygen-neon (ONe) novae that can occur at the surface of white dwarfs whose masses
are larger than $1.25{\rm M}_{\odot}$ (BT24). The yields from  Jos\'e \& Hernanz (1998)
are adopted for ONe novae for all metallicities, though their yields are for the solar metallicity.
This is because chemical yields of ONe novae are  only available for the solar metallicity (BT24).
Since it is not well known how gas-phase P and Cl can be accreted onto dust grains,
we do not consider dust growth through accretion of gas-phase P and Cl on dust grains.
Therefore, we need to assume the ratios of gas- to dust-phase abundances for P and Cl
in calculating interstellar chemistry related P- and Cl-bearing molecules, as later described.
It should be noted here that gas- and dust-phase abundances of all other elements observed
in the local diffuse interstellar medium are used for the present astrochemistry calculations too.

\begin{table}
\centering
\begin{minipage}{85mm}
\caption{ A summary for input parameters from galaxy-scale simulations
that can be used for astrochemistry calculations of molecular abundances
in the postprocessing of simulation data. }
\begin{tabular}{ll }
{Input parameters} &  Definition \\
 $\rho_{\rm g}$ & Gas density \\
 $T_{\rm g}$ & Gas temperature\\
 $f_{\rm H_2}$ & Molecular fraction\\
 $f_{\rm d}$ & Dust mass fraction\\
 $T_{\rm dust}$ & Dust temperature \\
 $a_{\rm d}$ & Dust size \\
 $A_{\rm V}$ & Dust extinction \\
 $f_{\rm g, \it i}$  & Gas-phase abundance ($i=$C, Na, P, Cl etc)\\
 $F_{\rm UV}$ &  UV radiation field strength \\
 $\zeta_{\rm CR}$  & Cosmic ray ionisation rate \\
\end{tabular}
\end{minipage}
\end{table}

\subsection{Dust model}

Since the models for 
the formation, growth, and destruction processes of dust in ISM,
photoelectric heating of ISM by dust,
${\rm H_2}$ formation on dust grains
in galaxies are essentially the same as those adopted in B13 and B15a,
we describe these very briefly.
The present dust model consists of (i) dust formation through
SNe Ia and SNe II and AGB stars,  (ii) dust growth due to accretion
of gas-phase metals on dust grains,
and (iii) dust destruction by core collapse supernovae (CCSNe) and SNe Ia.
Chemical abundances of 13 elements,  H, He, C, N, O, Na, Mg, Si, P, S, Cl, Ca, and Fe
can be investigated in the present simulations, and gas-phase and dust-phase
abundances are input to interstellar chemistry calculations to predict
the abundances of interstellar molecules.

Chemical yields for CCSNe, SNIa, and AGB stars are assumed to be the same as those
adopted in B13.
The mass
fraction of metals that are locked up in dust grains 
(``dust condensation efficiency'')  for each chemical component
from each  stellar type is assumed to be the same adopted in B13 too.
Dust growth by accretion of metals of ISM onto preexisting cores 
is parameterised by ``dust accretion timescale'' denoted as $\tau_{\rm acc}$ (B13).
The simulation code in B13 and B15a can investigate both the models with
fixed $\tau_{\rm acc}$ and variable one depending on local gas density and temperatures
of SPH particles.
We here adopt a fixed  $\tau_{\rm acc}=2.5 \times  10^8$ yr for all dust components,
because our previous simulations demonstrated
that  the models with $\tau_{\rm acc}=2.5 \times  10^8$ yr  can best reproduce 
the present-day dust abundances of MW-type disk galaxies among models with
different $\tau_{\rm acc}$ (B13).
Dust destruction though supernova blast waves
in the ISM of galaxies
can be parameterised by the dust destruction time scale
denoted as $\tau_{\rm dest}$.
Since B13 already found that 
simulations with $\tau_{\rm dest}\approx 2\tau_{\rm acc}$ can better reproduce
various dust and gas properties of galaxies,
we here adopt 
$\tau_{\rm dest}=5.0 \times  10^8$ yr for all dust components.

Photoelectric heating of ISM by dust has been demonstrated to influence
not only the thermodynamics of ISM but also galaxy-wide star formation
(e.g., B15a; Osman et al. 2022). We however do not incorporate such heating effects
in the present simulations, because additional parameters for the heating processes
need to be assumed in the simulations.
Molecular hydrogen is assumed to be formed on dust grains,
and the mass fraction ($f_{\rm H_2}$) for each particle
at each time step is estimated by the model adopted in B13 (see 2.3 of B13 for the
details):
$f_{\rm H_2}$ is determined by local far-UV radiation fields, gas densities,
and dust abundances. 
In these dust modelling, we do not consider 
dust coagulation and dust size evolution in ISM,
though such dust processes could influence the time evolution of dust abundances
and ${\rm H_2}$ formation on dust grains: these processes will be appropriately incorporated 
in our future simulations.

We need to derive  dust temperature ($T_{\rm dust}$) for each gas particle
in a simulation
in order to calculate the abundances of interstellar molecules using
the astrochemistry code adopted in the present study (F15).
We derive
$T_{\rm dust}$ for a gas particle
based on the strength of ISRF ($U$) around each gas particle as 
follows:
\begin{equation}
T_{\rm dust}=16.4 f_{\rm Si} (U/U_0)^{1/6} + 22.3 f_{\rm C} (U/U_0)^{1/6},
\end{equation}
where $U$ are the strength of local ISRF around the gas particle,
$U_0$ is the local $U$ of the Galaxy ($1.05 \times 10^{-12}$ erg cm$^{-3}$;
Mathis et al. 1983),
and the mass fractions of carbon ($f_{\rm c}$)
and silicate dust ($f_{\rm Si}$)  are set to be 0.34
and 0.66, respectively.
This analytic formula was originally 
proposed by Draine (2009) and
the typical dust size of $0.1\mu$m is adopted in the present study.
The initial dust-to-metal-ratio
of gas disk is set to be 0.365 in all models of the present study.

\begin{figure*}
\psfig{file=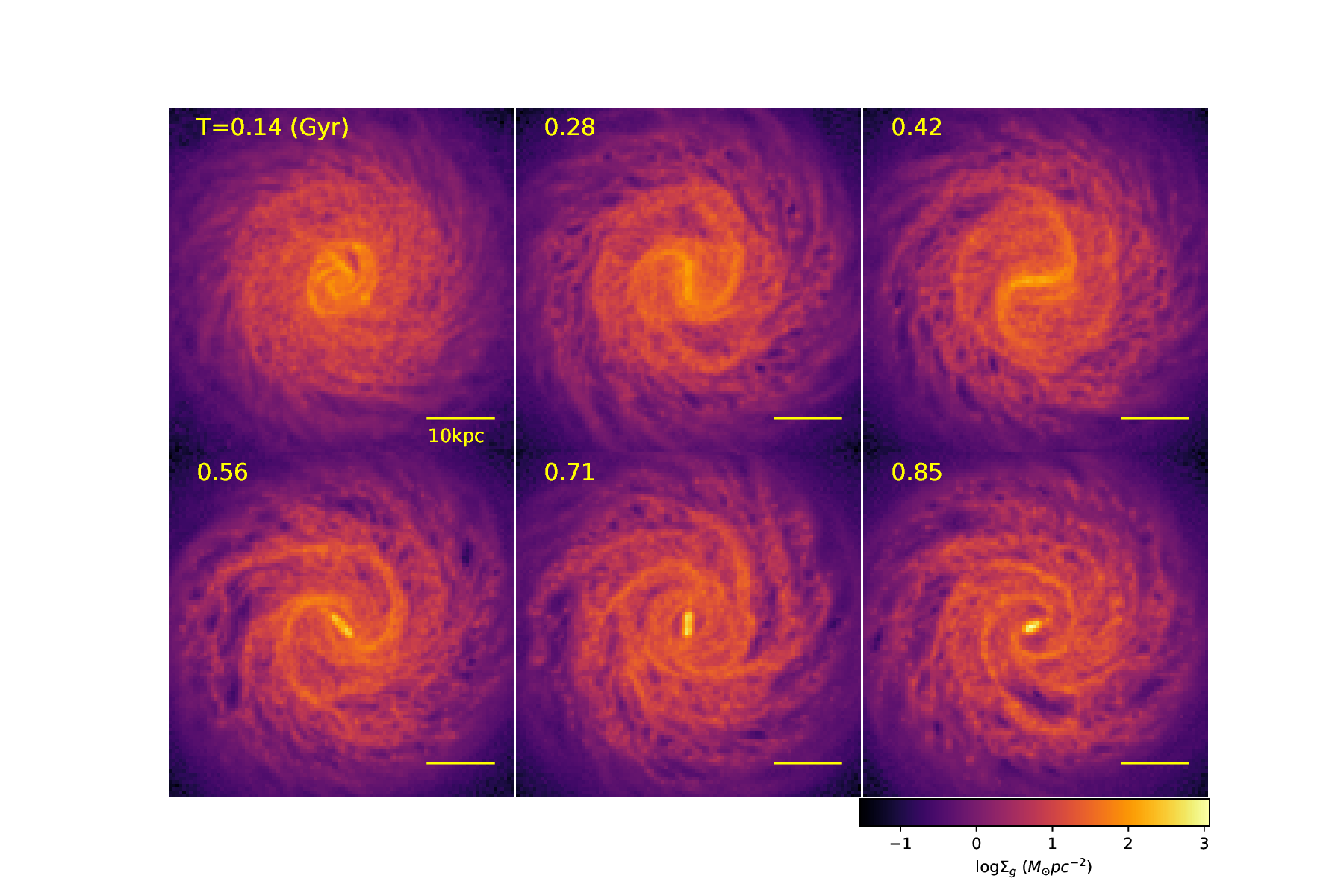,width=19.0cm}
\caption{
Time evolution of the surface gas densities ($\Sigma_{\rm g}$) projected onto the $x$-$y$
plane in the fiducial model. The gas disk is divided into $100 \times 100$ mesh points
(i.e., $3.5 \times 3.5$ kpc) to estimate $\Sigma_{\rm g}$ at each point.
}
\label{Figure. 2}
\end{figure*}

\subsection{ISRF and cosmic ray}

Each stellar particle has an age and a metallicity so that
the spectral energy distribution (SED) can be derived by using 
the stellar population synthesis code developed by Bruzual \& Charlot (2003)
for a given IMF. The local ISRF around a gas  particle is estimated by 
integrating SEDs from all stellar particles that are within the SPH smoothing length of the 
gas particle. We estimate the optical dust extinction ($A_{\rm V}$) of a gas particle
with the dust-to-gas-ratio of 0.0064 ($f_d$) using
the following formula (B15a):
\begin{equation}
\frac{ N_{\rm H} }{ A_{\rm V} }=
1.8 \times 10^{21} \;  {\rm atoms \; cm^{-2} \; mag^{-1}  },
\end{equation}
where $N_{\rm H}$ is the column density of hydrogen atoms, which can be estimated 
from the mass densities of gaseous particles along the line of sight.
Therefore, $A_{\rm v}$ depends both on local gas densities and $f_{\rm d}$.
In deriving dust-corrected local far-UV (FUV) flux from local ISRF,
we consider that the ratio of $A_{\rm FUV}$ to $A_{\rm V}$
is fixed at 2.56 (B15a).
We use the following normalised UV flux ($R_{\rm UV}$) to discuss
the spatial distribution of UV flux in a galaxy;
\begin{equation}
R_{\rm UV}=\frac{ F_{\rm UV} }{ F_{\rm UV,0} },
\end{equation}
where $F_{\rm UV}$ and $F_{\rm UV,0}$ are the local UV flux for a gas particle
and the local UV strength.

Interaction between ISM  and cosmic-ray (`CR') can influence the thermodynamics of
typical ISM in galaxies with
$n \sim 1$ atom cm$^{-3}$ and SFR$\sim 1$ ${\rm M}_{\odot}$ yr$^{-1}$ (B15).
However, we do not include this influence of CR in the present simulations,
mainly because introduction of additional parameters for CR heating can possibly
make it (unnecessarily) more complicated for the present study to physically interpret
the results of interstellar molecular abundances.
where $\zeta_{\rm CR}$ is total  cosmic ionisation rate.
If cosmic ray originates mainly from CCSNe and SNe Ia, then it is reasonable
for us to assume that the total flux of comic ray in a galaxy  is proportional
to the SFR of the galaxy. We therefore adopt the following
formula for the cosmic ionisation rate 
($\zeta_{\rm CR}$)
depending only on SFR: 
\begin{equation}
\zeta_{\rm CR} = 2 \times 10^{-16} (\frac{ {\rm SFR} }
{ {\rm 1.3 {\rm M}_{\odot} yr^{-1} } }) {\rm s}^{-1},
\end{equation}
where the reference SFR of $1.3 {\rm M}_{\odot}$ yr$^{-1}$ corresponds
to the present SFR of the Galaxy (e.g., Draine 2009).
We estimate the total SFR of a galaxy every $10^7$ yr in order to derive
$\zeta_{\rm CR}$ that is required for interstellar chemistry calculations later described.

\begin{figure*}
\psfig{file=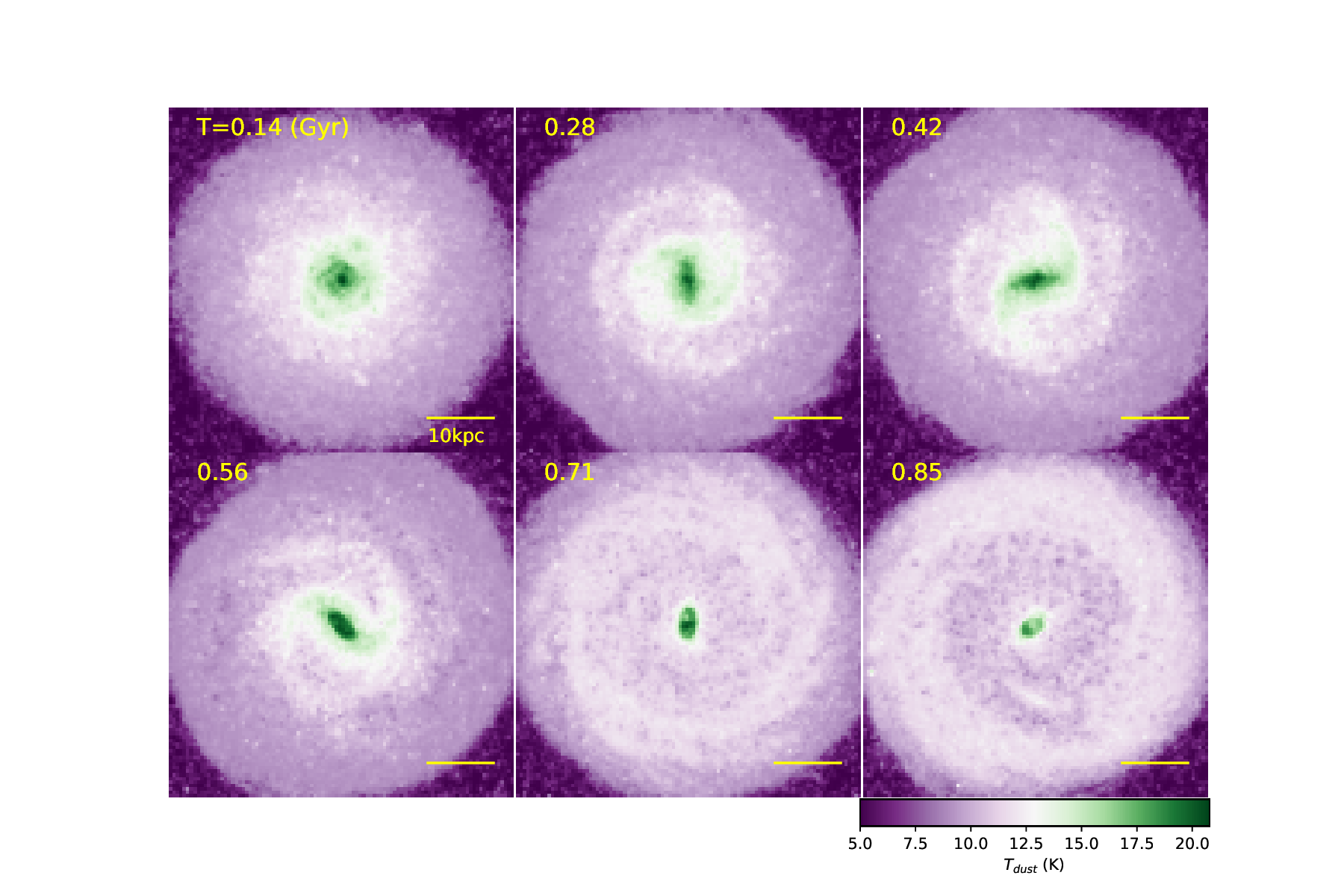,width=19.0cm}
\caption{
The same as Fig. 1 but for dust temperature, $T_{\rm dust}$.
}
\label{Figure. 3}
\end{figure*}

\subsection{Interstellar chemistry}

After we complete a galaxy-scale simulation,
we apply  our astrochemistry code (F15) to each gas partice in the simulation
in order to derive the abundances of
gas- and ice-phase molecules for the particle:
The input parameters (e.g., UV radiation fields)
for F15 in this postprocessing are stored for each gas
particle at selected time steps in the simulation.
In this postprocessing method, we assume that gas- and ice-phase molecules
do not influence galaxy evolution (e.g., dust and metal abundacne evolution).
Since the present simulations with the spatial resolution of $\sim 100$pc 
cannot resolve the pc-scale molecular cores with gas densities of $10^5$ atom cm$^{-3}$,
we use the physical parameters of high-density
gas particles (corresponding to giant molecular clouds)
for the calculations of molecular compositions within molecular cores.
We accordingly assume that all simulated GMCs have molecular cores with
a similar density in the present study, though there could be a variation in the core densities.

We utilise the gas-ice astrochemical code based on the rate equation approach (Rokko code; 
F15). 
We adopt a three-phase model, where the gas phase, a surface of ice, and the chemically inert bulk ice mantle, which covers dust grain cores, are considered (Hasegawa \& Herbst 1993). 
Around 100 monolayers of ice can be formed on grains, and the top four ice layers are assumed to be the surface, while the rest is considered as the bulk ice mantle  (Vasyunin \& Herbst 2013). 
Our gas-ice chemical network is based on that of Garrod (2013);
gas phase reactions, the interaction between gas and ice, and grain surface chemistry are considered.
The network includes 667 gas-phase species, 307 icy species, and $\sim$12,000 reactions in total.
The binding energy distribution, which is one of the most important parameters for describing the grain surface chemistry, is taken into account by the method developed in Furuya (2024). 
The binding energy of each species is assumed to follow the Gaussian distribution.
The mean value of the binding energy of each species is taken from Garrod (2013)
 and Wakelam et al. (2017), while the standard distribution is assumed to 
be 0.2 times the mean value.
Our chemical model includes both thermal and non-thermal desorption. Non-thermal desorption mechanisms in our model include photodesorption, desorption by stochastic heating by cosmic rays, and reactive desorption.
More details can be found in Furuya et al. (2015) and Furuya (2024).

It is assumed that (i)
 the cores have a H$_2$ density of 10$^4$ cm$^{-3}$ and 
(ii) the visual extinction of is fixed at 10 mag: the external UV radiation is therefore negligible.
The CR ionisation rate of H$_2$ is set to $2\times10^{-16}$ s$^{-1}$.
We follow the molecular evolution for 10$^{6}$ yr which is larger than the freeze-out timescale of gas-phase molecules onto dust grains (i.e., the timescale of ice formation).
Elemental abundances and dust temperatures are taken from the Galaxy simulations.
As our chemical network considers volatile (gas and ice) components only, the direct use of the predicted elemental abundances would not be appropriate; it has been known that chemical network simulations fail to reproduce the molecular line observations of molecular clouds, if one adopts gas-phase elemental abundances (or depletion factors) observed in diffuse clouds. 
To reproduce the molecular line observations, smaller gas-phase elemental abundances (i.e., larger depletion factors) than those observed in diffuse clouds are required (so-called low metal abundances; e.g., Graedel et al. 1982).
In our chemical network simulations, we define the depletion factor for each element as the ratio of low-metal abundance to that of the solar abundance.
In the simulations of each molecular cloud core, we use the elemental abundances taken from the Galaxy simulations multiplied by the depletion factor defined above.
This means that it is assumed that the depletion factor is the same for all cores in the Galaxy.
The input parameters from galaxy-scale simulations for the code Rokko are summarised in Table 2,
and some of them are fixed (e.g., $\rho_{\rm g}=10^4$ cm$^{-3}$) due to the lack of spatial resolutions
of the simulations).
Ice on dust grains can possibly suppress dust growth (e.g., Ferrara et al. 2016), however, we do not
include this ``ice feedback'' effect in the present simulations. We discuss this later in \S 4.
The present study 
does not discuss galaxy-scale distributions of
molecular species formed in 
{\it circumstellar environments of evolved stars and planetary 
nebulae }
(e.g., Ziurys et al. 2018).
We will discuss 
interplays between astrochemical processes and galactic dynamics
suggested by recent observations (Mendoza et al. 2021) in our future papers.

\begin{figure*}
\psfig{file=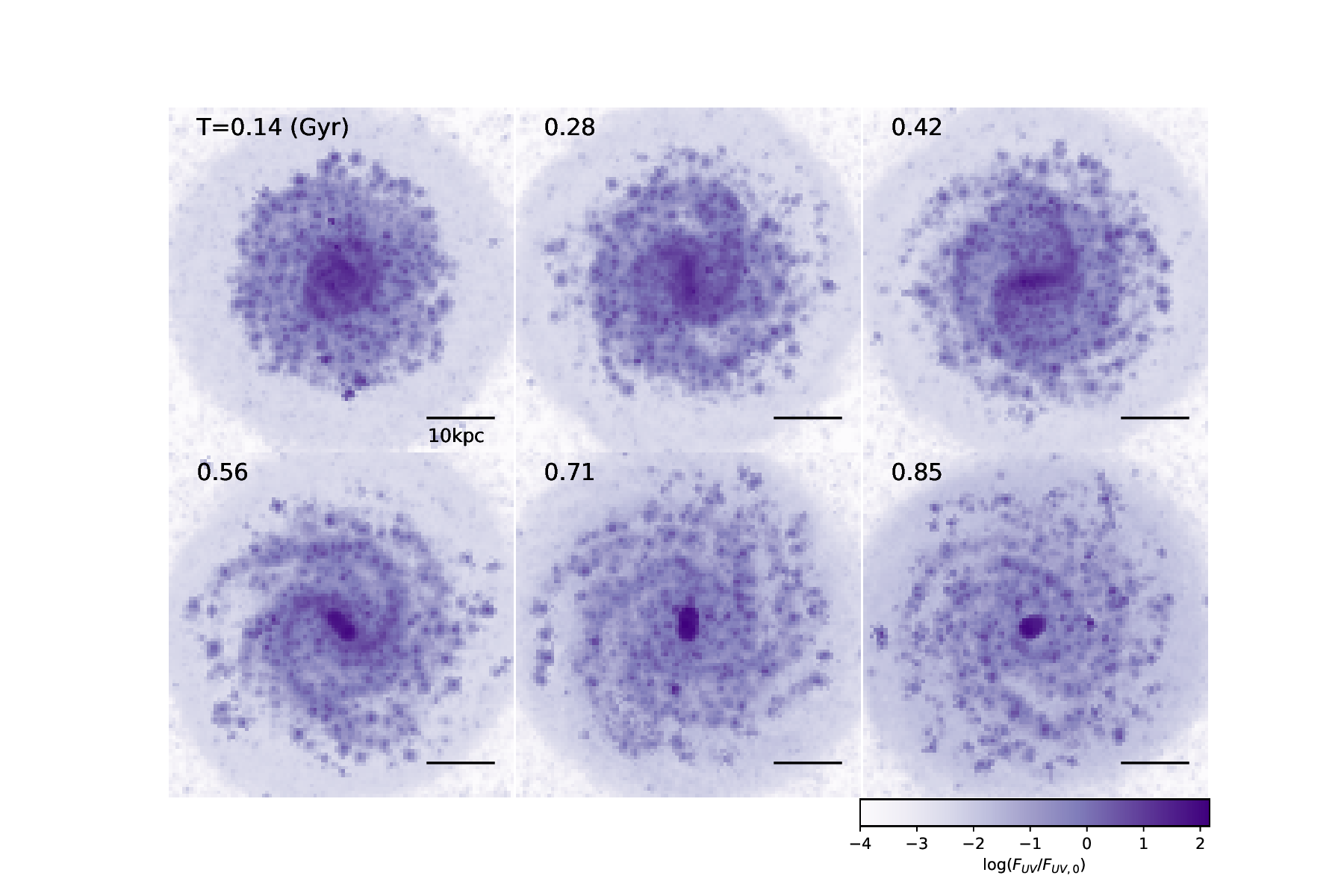,width=19.0cm}
\caption{
The same as Fig. 1 but for normalised UN radiation fields, $R_{\rm UV}$,
where $R_{\rm UV}=F_{\rm UV}/F_{\rm UV,0}$.
}
\label{Figure. 4}
\end{figure*}

\subsection{Analysis of selected particles}

It takes 3 minutes to several hours for the adopted interstellar 
chemistry models (F15) to calculate $\approx 10^6$ yr evolution of 
$\approx 1000$ molecular species for just one  gas particle in a galaxy.
Therefore, it is practically infeasible for the present study to
derive the abundances of interstellar molecules for all ($N\approx 10^6$) gas
particles at each time step  in a simulation.
We accordingly need to select particles ($N<1000$) in a galaxy and thereby
investigate their abundances of interstellar molecules.
To do so, we first select gas particles from a simulated galaxy for  each of ten radial bins 
($0 \le R/R_{\rm s} \le 1$) and calculate the time evolution of 
molecular species in the selected particles  for $10^6$ yr. This calculation is
done using the physical parameters of the selected particles
only after entire simulations have been completed. This ``post-processing'' 
is the best way to derive the abundances of interstellar molecules in this preliminary
study of molecular evolution in galaxies, though
the time evolution of molecular abundances can possibly influence galaxy evolution.
It is our future study to incorporate both (i)  the formation and evolution of interstellar
molecules in galaxies and (ii) their influences on galaxy evolution in galaxy-scale simulation
self-consistently.

Physical conditions of local ISM across a galaxy can be quite different due to diverse
star formation activities within molecular clouds, 
dust-to-gas-ratios,  metallicities, and ISRF strengths. 
We particularly focus on how the present results depend on 
 $T_{\rm dust}$ and $f_{i, g}$ ($i=$ C, N, O etc; gas-phase chemical abundances),
because  $T_{\rm dust}$ and $f_{i, g}$ are found to be different among gas clouds
at different positions and thus
are particularly important in discussing the possibly diverse molecular abundances.
In order to understand how these diversities influence the abundances of interstellar
molecules, we take the following three steps.
First we derive the means and 1$\sigma$ dispersions of all input parameters
for interstellar chemistry calculations
at each radial bin in a simulated galaxy at a given time step using all particles in each bin.
Second, using these mean values,
we calculate the molecular abundances in each radial bin to derive the radial gradients of 
the abundances.
Third, we change the values of $T_{\rm dust}$ and $f_{i, \rm g}$ only (other 
input parameters being fixed) and run 100 different models for each radial bin.
By doing so, we can better understand how diverse $T_{\rm dust}$ and $f_{i, \rm g}$ can
influence the molecular diversity in the simulated galaxy.
In this third step, 
we randomly select $T_{\rm dust}$ or $f_{i, \rm g}$ or both by using a random number generator
following the Gaussian distribution function with the mean and dispersion being
derived from each radial bin in a simulation.
For example, a new dust temperature ($T_{\rm dust}^{'}$) in a newly generated set of
input parameters for a radial bin  is as follows;
\begin{equation}
T_{\rm dust}^{'} = T_{\rm dust,m} + \delta T_{\rm dust},
\end{equation}
where $T_{\rm dust, m}$ is the mean  dust temperature at the radial bin and
$\delta T_{\rm dust}$ is randomly generated based on the 1$\sigma$ dispersion of $T_{\rm dust}$.
Likewise, a new gas-phase abundance is generated as follows:
\begin{equation}
f_{i, \rm g}^{'} = f_{i, \rm g,m} + \delta f_{i, \rm g},
\end{equation}
where $f_{i, g, m}$ is the mean $f_{i, \rm g}$ at the radial bin. We randomly change gas-phase abundances
of all 13 elements that are necessary for astrochemistry calculations.

\begin{figure}
\psfig{file=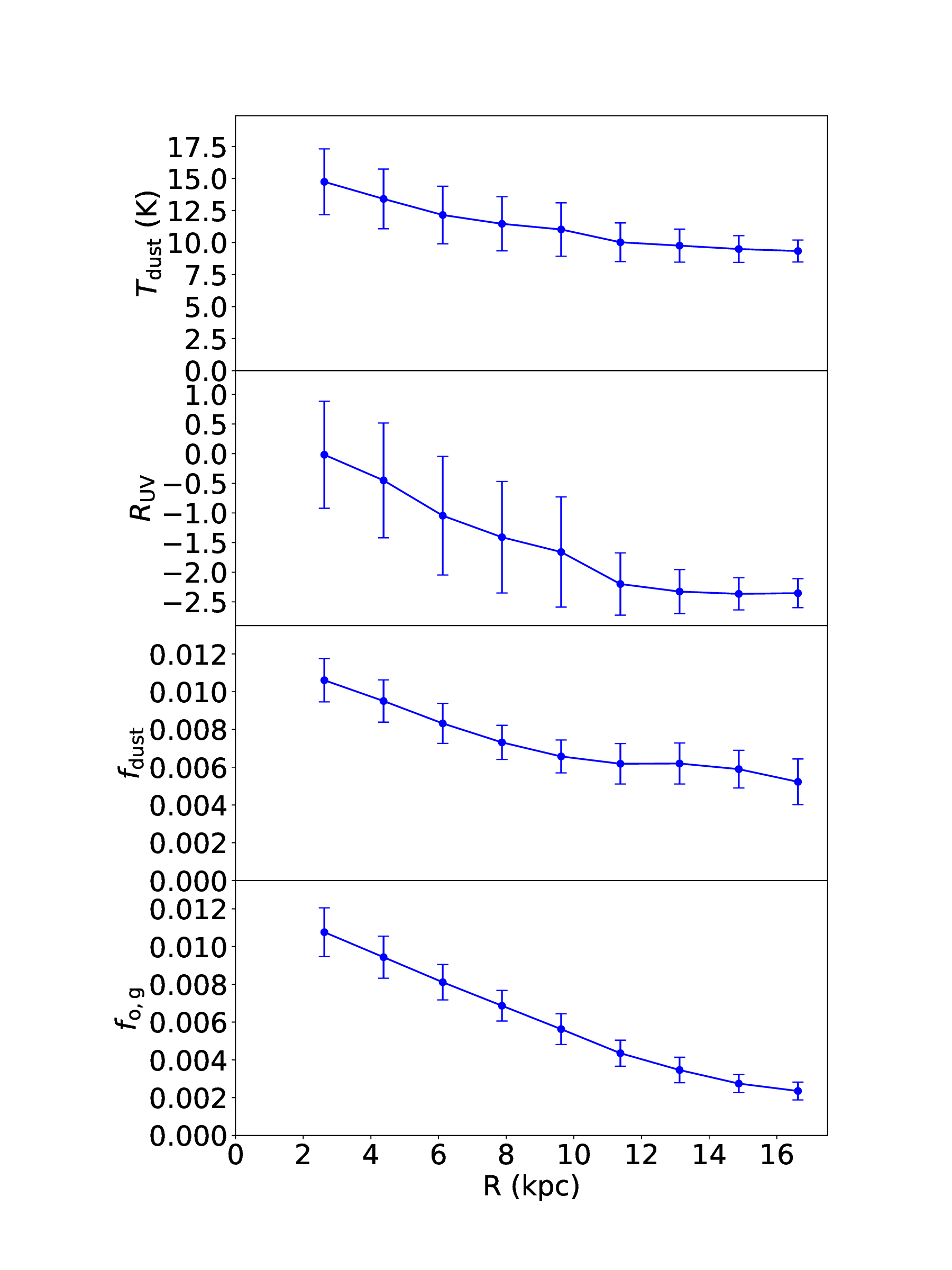,width=8.5cm}
\caption{
Radial profiles of $T_{\rm dust}$ (top),
$\log R_{\rm UV}$ (second from the top),
$f_{\rm dust}$ (second from the bottom), 
and $f_{\rm o,g}$ (bottom) 
in the fiducial model.
The error bars at each radial bin represent $1\sigma$ dispersions of these physical properties.
These properties for the central bulge ($R<2$ kpc) in the first bin
are estimated using all particles within $R<2$ kpc.
}
\label{Figure. 5}
\end{figure}
\begin{figure}
\psfig{file=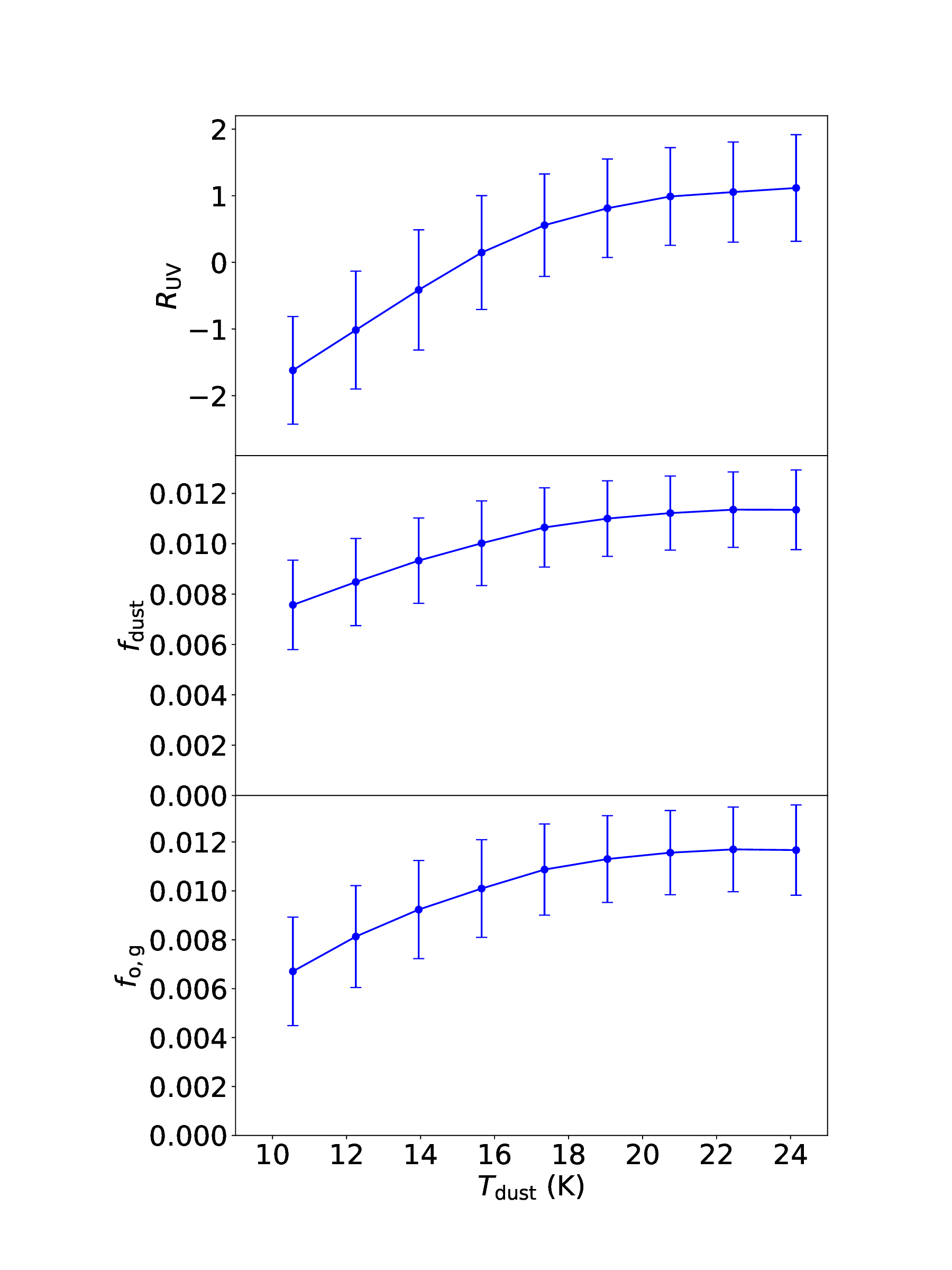,width=8.5cm}
\caption{
Correlations of $\log R_{\rm UV}$ (top),
$f_{\rm dust}$ (middle), and $f_{\rm o,g}$ (bottom) 
with $T_{\rm dust}$  in the fiducial model.
The error bars at each $T_{\rm dust}$ bin represent  $1\sigma$ dispersions of these physical properties.
}
\label{Figure. 6}
\end{figure}

\section{Results}

\subsection{Global evolution of ISM:Gas density, dust temperature, and ISRF distributions}

Figs. 2, 3, and 4 describe the time evolution of the spatial distributions of
gas surface densities ($\Sigma_{\rm g}$), dust temperatures ($T_{\rm dust}$),
and normalised UV strengths ($R_{\rm UV}$), respectively,  during
$0.85$ Gyr dynamical evolution of the fiducial disk galaxy model.
A strong stellar bar can quickly form in the inner region of the galaxy
from global bar instability
within $\approx 0.3$ Gyr due to the adopted smaller bulge mass
fraction (0.17) and the higher baryonic mass fraction ($\approx 0.4$) within $R_{\rm s}$.
The bar can generate a number of spiral-arm structures, where high-density
gaseous clumps can form and subsequently are converted into new stars.
These new stars are the major sources of UV radiation that can influence 
(i) gas-gas and gas-dust  interstellar chemistry of ISM and (ii) $T_{\rm dust}$
evolution in this model.
Given that UV-radiation is almost completely  shut off (${\rm A_{\rm v}=10}$ mag) in the present 
astrochemistry calculations,
such strong radiation can only influence $T_{\rm dust}$, which can influence dust-phase astrochemistry.
As shown in the 2D $\Sigma_{\rm g}$ map of gas density,
CCSNe from these new stars generates low-density hole-like regions
due to their energetic thermal and kinetic feedback effects.
2D spatial distributions of gas-phase metallicities (e.g., oxygen abundances,
$\Sigma_{\rm o, g}$) and dust mass ($\Sigma_{\rm dust}$)
are very similar to those of $\Sigma_{\rm g}$ in Fig. 2 (See the Appendix A for 
the 2D distributions of $\Sigma_{\rm dust}$ and $\Sigma_{\rm o, g}$).

In comparison with $\Sigma_{\rm g}$, 
2D distributions of $T_{\rm dust}$ do not show  spiral-arm structures so clearly,
though the central bar-like structure can be seen in the early dynamical evolution
of the disk ($T=0.28$, 0.42 and 0.56 Gyr).
This is mainly because $T_{\rm dust}$ of a local region is determined by the total strength of ISRF 
from all stars (not just young stars formed in spiral arms).
Clearly, $T_{\rm dust}$ is significantly higher  along the central stellar bar,
where the higher surface mass density of old and new stars causes the stronger ISRF.
Such higher $T_{\rm dust}$ along the stellar bar indicates that interstellar chemistry
could be significantly different between barred and unbarred regions in a disk galaxy.
A strong radial gradient of $T_{\rm dust}$ can be also seen in the disk,
which is consistent with recent observational results of nearby disk galaxies 
(e.g., Groves et al. 2012 for M31). Since $T_{\rm dust}$ is a key parameter
in interstellar chemistry, higher $T_{\rm dust}$ in the inner regions 
of a  galaxy can result in the strong radial gradients of molecular
abundance within  the galaxy.
The sharp edge of the simulated $T_{\rm dust}$ map is due to the assumed sudden
truncation of the old stellar disk in this model.

As shown in Fig. 4,
there are numerous small clumps with sizes less than a few hundreds pc
in the 2D map of $R_{\rm UV}$ of the disk galaxy at all time steps.
Each of these clumps corresponds to
the location of a high-density molecular gas with active star formation,
where UV radiation from young stars is rather strong. 
The sizes of these clumps vary significantly, which reflects the diverse masses and sizes of
molecular gas clouds formed in the gas disk.
The central stellar bar region has rather high $R_{\rm UV}$ due to the formation of new stars 
within the stellar bar ($T=0.25$, 0.42, and 0.56 Gyr).
This high $R_{\rm UV}$ and high $T_{\rm dust}$ within the central bar suggests that
interstellar chemistry of cold gas clouds  could be different between
within and outside stellar bars in disk galaxies.
It would be an important observational study to investigate
whether the compositions  of interstellar ice 
(e.g., ${\rm H_2O}$-to-${\rm CO_2}$-ice-ratios) can be different between
gas clouds within and outside stellar bars in disk galaxies.

\begin{figure}
\psfig{file=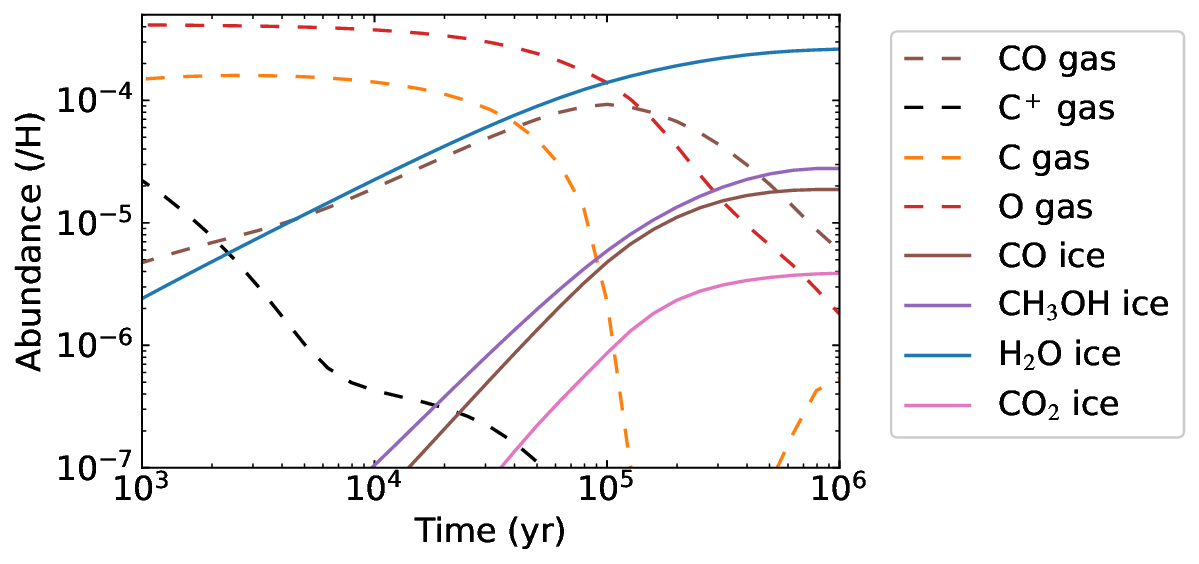,width=8.5cm}
\caption{
Time evolution of gas-phase CO (blue),  ${\rm H_2O}$ ice (orange), ${\rm CO_2}$ ice (green), 
and ${\rm CH_3OH}$ ice (red)
for a selected gas particle with $R=8.5$ kpc, $T_{\rm dust}=11.5$ K,
and $f_{\rm H_2}=0.89$  at $T=0.28$ Gyr in the fiducial model.
}
\label{Figure. 7}
\end{figure}

\begin{figure}
\psfig{file=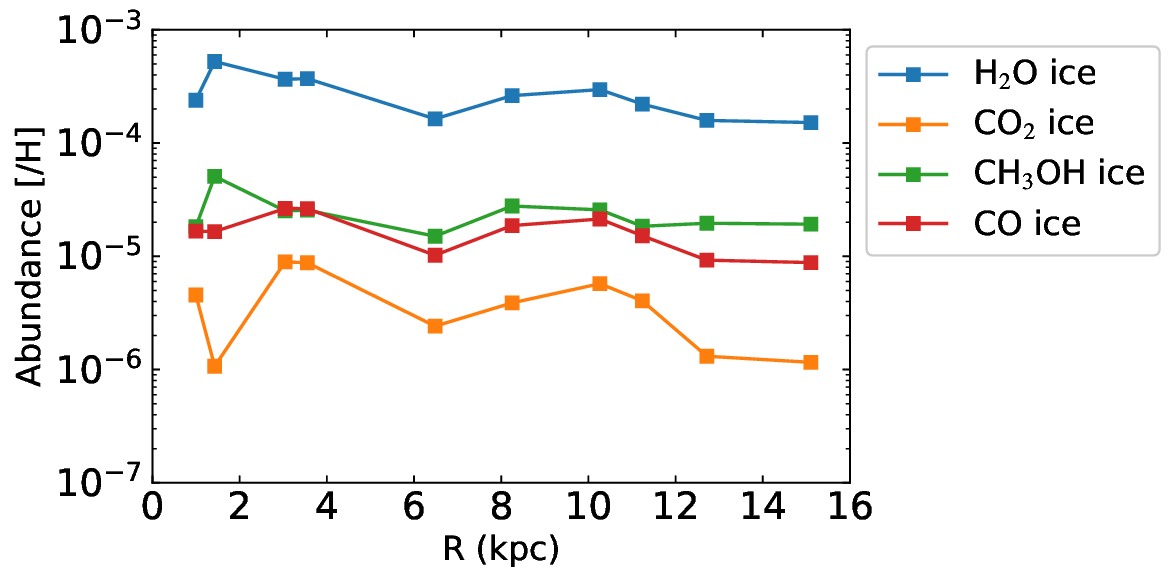,width=8.5cm}
\caption{
Radial profiles of
CO ice (blue),  ${\rm H_2O}$ ice (orange), ${\rm CO_2}$ ice (green), 
and ${\rm CH_3OH}$ ice abundances (red) with respect to hydrogen within the stellar
disk ($R\le R_{\rm s}$) at $T=0.28$ Gyr in
the fiducial model. At each radial bin, one gas particle with $10 \le T_{\rm dust} \le 12$ (K)
and $f_{\rm H_2}\ge 0.8$ is randomly selected and plotted in this figure.
}
\label{Figure. 8}
\end{figure}

\begin{figure*}
\psfig{file=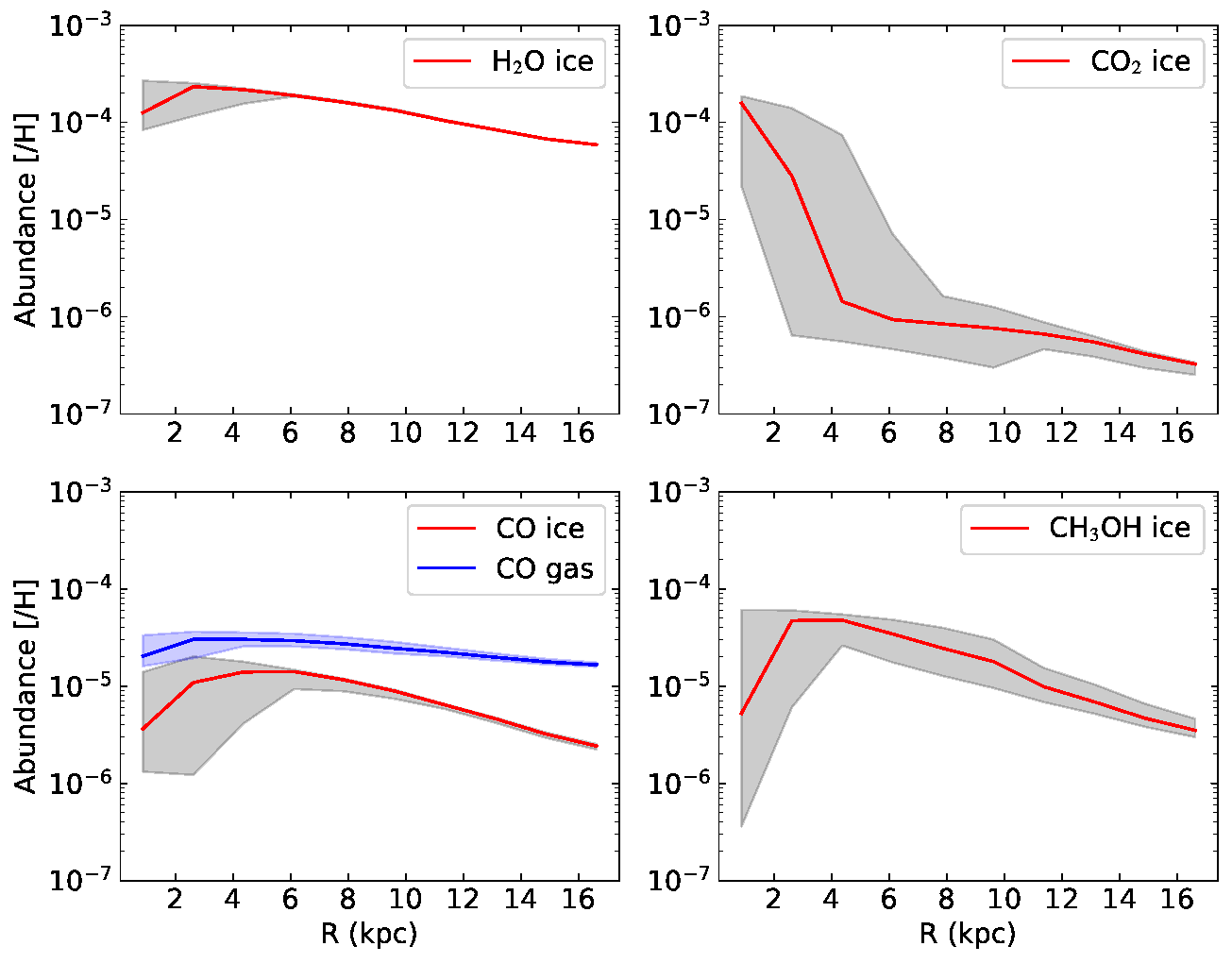,width=18.cm}
\caption{
Radial profiles of ${\rm H_2O}$ (upper left),
${\rm CO_2}$ (upper right), ${\rm CO}$ (lower left), and ${\rm CH_3OH}$ ice
in the simulated galaxy of the fiducial model  at $T=0.28$.
The red line denotes the radial profiles of these ice abundances
estimated from the mean properties (e.g., $T_{\rm dust}$) of
gas particle at each radial bin.
The black-coloured areas indicates the dispersions
of the ice abundances, however, the dispersion of each ice component
at each radial bin is not calculated from 
the ice abundances of all gas particles at the bin. Instead, the dispersion
is calculated by assuming that (i) there are 100 particles at a radial bin,
(ii) the mean and dispersion of $T_{\rm dust}$ of the 100 particles are the same as those
estimated from all gas particles at the bin,
(iii) $T_{\rm dust}$ of the 100 particles follows the Gaussian distribution,
and (iv) physical properties other than $T_{\rm dust}$ for the 100 particles are the same as the mean
values estimated from all gas particles at the bin. Therefore, this figure describes
how $T_{\rm dust}$ dispersions can influence the radial profiles of
the four ice abundances in galaxies. For example, the water ice abundance does not depend so strongly on
$T_{\rm dust}$ thus has a smaller dispersion:
the abundance is predominantly determined by gas-phase abundances of H and O.
The ${\rm CO_2}$ ice abundance is very sensitive to $T_{\rm dust}$ so that it can show
a steep radial decline following the negative radial gradient of $T_{\rm dust}$.
The simulated abundances ratios of ${\rm CO_2}$, CO, and ${\rm CH_3OH}$ to ${\rm H_2O}$ will be 
compared to the corresponding observations to assess the viability of the adopted models for
galaxy-scale molecular formation.
}
\label{Figure. 9}
\end{figure*}

These clear differences in the spatial distributions of $\Sigma_{\rm g}$, $T_{\rm dust}$,
and $R_{\rm UV}$, which are the three of key physical parameters for interstellar chemistry,
strongly suggest that gas-phase and dust-phase chemical processes are quite diverse
between gas clouds located at different regions of 
disk galaxies: see Appendix A for  differences and similarities
in the spatial distributions of these physical parameters
(e.g., dust surface densities).
One of ways to quantify these differences is to investigate the radial gradients of the means
and 1$\sigma$ dispersions of these physical parameters.
Fig. 5 shows that steep negative radial gradients
can be see in the radial profiles
of $T_{\rm dust}$, $R_{\rm UV}$, $f_{\rm dust}$, and $f_{\rm o, g}$ 
at $T=0.28$ Gyr in the disk model.
These results clearly demonstrate that there should be strong radial gradients of
gas- and ice-phase  molecules and ice in this disk galaxy.
Given that radial gradients of $R_{\rm UV}$ and $f_{\rm o, g}$ are quite steep,
interstellar ice species the formation processes of which strongly
depends on the two properties can possibly have the steep radial gradients as well.

Fig. 5 also shows that  1$\sigma$ dispersions of these key properties are quite significant even for a given
$R$, which physically means  that there are tangential variations in these properties.
The 1$\sigma$ dispersions also become smaller at larger $R$, where local star formation
rates are significantly lower. Such lower levels of  star formation activities result in
lower degrees of dust and metal enrichment and $T_{\rm dust}$ and $R_{\rm UV}$ variations
in the outer regions of the galaxy.
These results indicate that (i) ice abundances in molecular clouds with similar $R$ in a galaxy
can be quite different due to the possibly diverse $T_{\rm dust}$, $R_{\rm UV}$,
and gas-phase abundances of the clouds
and (ii) such differences could be not so remarkable in the outer regions of the galaxy: we discuss
these two points based on the simulated ice abundance distributions.

Fig. 6 describes the  physical correlations of $T_{\rm dust}$ with 
$R_{\rm UV}$, $f_{\rm dust}$, and $f_{\rm o, g}$ at $T=0.28$ Gyr in the disk model.
It is clear from this figure that gas clouds with higher $T_{\rm dust}$ 
are likely to have higher $R_{\rm UV}$, higher $f_{\rm dust}$, and higher $f_{\rm o, g}$.
This is mainly because star formation and the resultant enrichment of metals and dust
can proceed very efficiently in the disk's inner region where $T_{\rm dust}$ is higher
due to stronger ISRF (caused by the higher stellar density).
Dispersions of these 
$R_{\rm UV}$, $f_{\rm dust}$, and $f_{\rm o, g}$
for a given $T_{\rm dust}$ are larger, which means that interstellar
chemistry can be quite diverse even for gas clouds with similar  $T_{\rm dust}$.
Steep radial gradients of these physical properties
and correlation of these  shown in Figs. 5 and 6 for $T=0.28$ Gyr
are confirmed to be seen also for other time steps in this model.

\begin{figure*}
\psfig{file=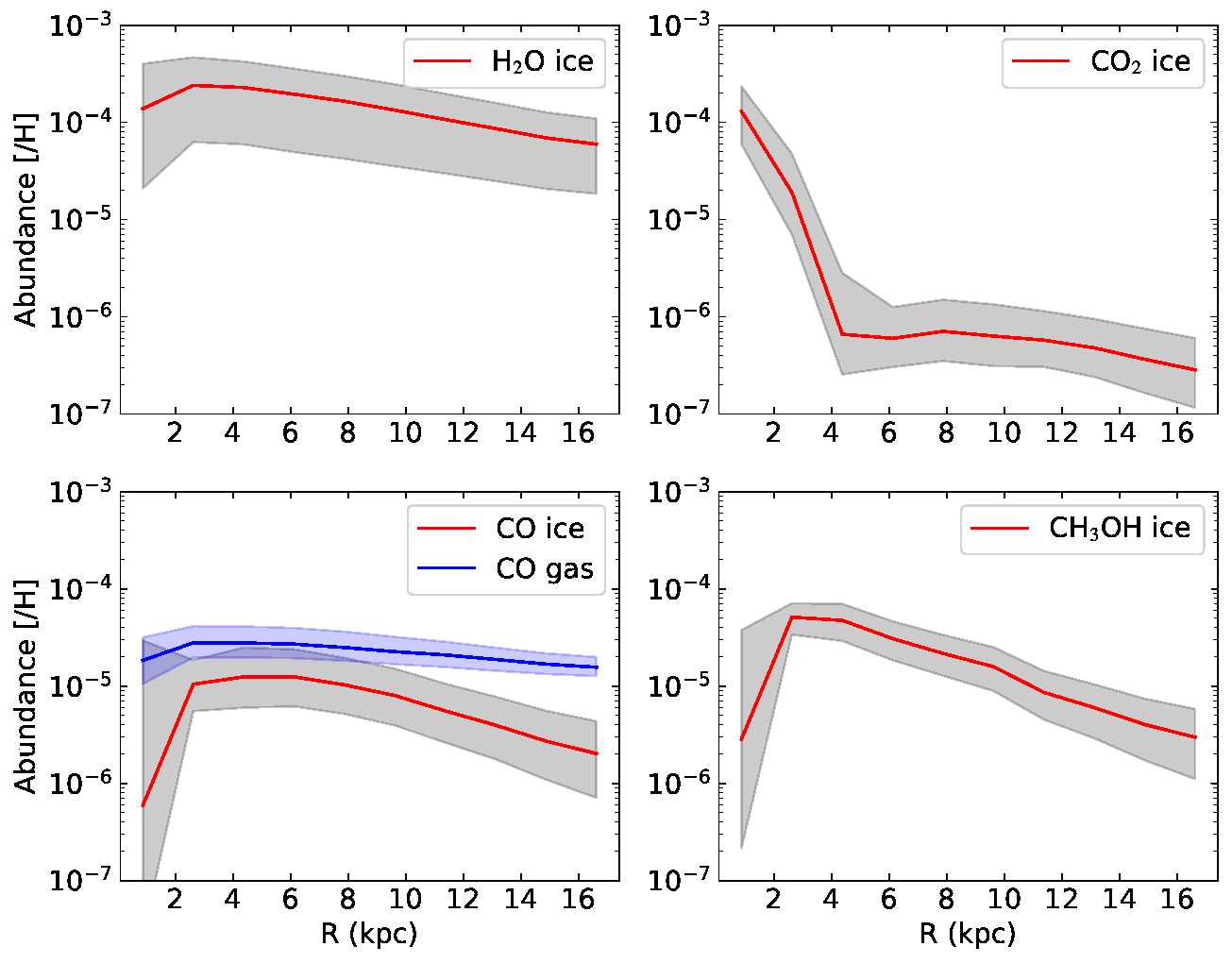,width=18cm}
\caption{
Same as Fig. 9 but for the dispersions of ice abundances caused by
dispersions in  gas-phase chemical abundances of all elements used for astrochemical calculations.
Accordingly, $T_{\rm dust}$ at each radial bin is fixed at the mean value estimated
from all gas particles at the bin.
}
\label{Figure. 10}
\end{figure*}

\subsection{Radial gradient of molecules}

Fig. 7 describes the time ($T$) evolution of various gas- and ice-phase
molecular abundances of a selected gas particle with $Z \approx Z_{\odot}$ 
and $f_{\rm H_2}=0.80$ (corresponding to a molecular cloud)
at $R\approx 8.5$ kpc in the fiducial model.
Here we calculate these abundances for $0\le T \le 10^6$ yr (rather than $10^7$ yr),
 and use the final values
of these abundances at $T=10^6$ yr to derive the galaxy-scale distributions of the abundances.
Although gas-phase abundances are still rapidly evolving at $T \approx 10^6$ yr,
ice components become very close to their maximum values at $T \approx 10^6$ yr. Given that
only the ice abundances are mainly analysed and discussed in this paper, 
adopting the ice abundances at a particular time step of  $T=10^6$ yr 
is highly unlikely to influence the main results of the present simulations.

${\rm H_2O}$ ice in this particular gas particle
is found to become the most abundant after  $T \approx 10^5$ yr in the present
astrochemistry calculations based on galaxy-scale simulations. This results  is
broadly  consistent with
previous works (e.g., F15) in which  initial parameters for chemistry calculations are
manually given.
The abundances of CO, ${\rm CH_3OH}$, ${\rm CO_2}$, and ${\rm NH_3}$ ice 
at $T=10^6$ yr derived from galaxy-scale simulations in the present study are also consistent 
with previous works (F15), which confirms that the present galaxy-scale
simulations (and possibly galaxy-scale simulations from future works by other groups)
can properly predict the ice abundances in molecular clouds.
It should be noted here that $A_{\rm V}$ for molecular cores  is assumed to be 10 mag for all gas
clouds in a simulated galaxy.
Given that $A_{\rm V}$ in real cores of molecular clouds can be different
depending on $Z$,
the adopted assumption might lead to over- or under-estimation of ice abundances within
molecular cores of  galaxies.

Fig. 8 shows the abundances ${\rm H_2O}$,  ${\rm CO_2}$, CO, and ${\rm CH_3OH}$ ice 
for randomly selected ten gas particles for each radial bin in this simulated galaxy.
Clearly,  there are no strong radial gradients of ${\rm H_2O}$,
${\rm CO_2}$, and ${\rm CH_3OH}$ ice in this simulated galaxy,
{\it if the gas particles with high $f_{\rm H_2}$  are randomly selected.}
This is essentially because  the physical properties of ISM (e.g., dust-to-gas-ratios and 
$T_{\rm dust}$) are quite different in different gas clouds. Nevertheless,
The abundance of ${\rm H_2O}$ ice is always by more than a factor of $\sim 10$
 higher than those of other ice components,
which implies that the mass fraction of ${\rm H_2O}$ ice within molecular
cores does not vary so much with galactic radii.
The radial variations of ${\rm CO_2}$ and  CO ice
are more remarkable compare to other two ice components in this galaxy.
This is probably because the formation of ${\rm CO_2}$ and  CO  ice
is more sensitive to chemical abundances and dust properties of local ISM.

We are unable to compare
these derived negative gradients of ice abundances over the entire gaseous disk of 
a disk galaxy with corresponding observations due to the lack
of observational studies on the radial gradients. 
However, Fontani et al. (2022) have
recently discovered an almost flat  radial abundance gradient of gas-phase ${\rm CH_3OH}$ in the outer
part of the Galaxy (13 kpc$<R<$ 19 kpc). The simulated radial gradient of ${\rm CH_3OH}$ ice
is clearly negative due largely to the adopted  initial negative metallicity
gradient.
Although their results are for gas-phase (not ice-phase) abundances,
this inconsistency between the observation and the present study implies either that 
(i) the metallicity gradient in the very outer part of the Galaxy is not so steep as assumed in the present study, or 
(ii) the desorption efficiency of the CH3OH ice or the destruction efficiency of the CH3OH gas is different in the outer Galaxy, 
or (iii) the present study does not so well grasp some essential ingredients that can enhance the 
formation of CH3OH ice in low-metallicity environments. 
Given that such a flat radial gradient can be seen also in HCO and ${\rm H_2CO}$ (Fontani et al. 2022),
we will need to discuss the physical origins of this inconsistency in detail in our forthcoming
paper.
Furthermore,
Fontani et al. (2024) have recently investigated the abundances of various organic
species within star-forming regions of the very 
outer part of the Galaxy ($R \approx 23$ kpc) and thereby compared the results
with their chemical models. For example, they have found that
${\rm CH_3OH}$ abundances are consistent with those of the organic-poor cores
discovered in the LMC.  Our simulations also have shown that the outer metal-poor
regions of the Galaxy have significantly low ${\rm CH_3OH}$ abundances, which is qualitatively
consistent with the above result by Fontani et al. (2024).

\begin{figure*}
\psfig{file=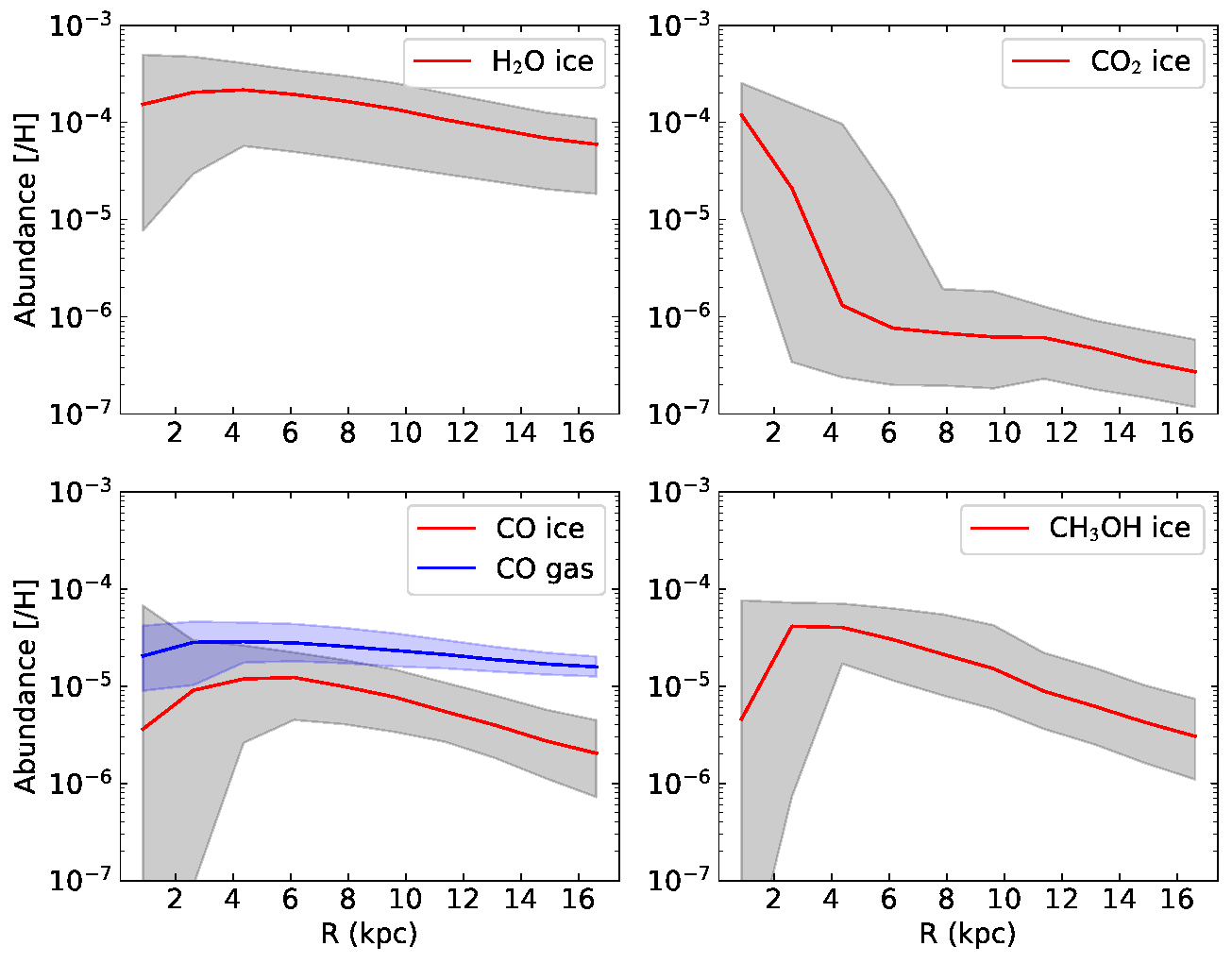,width=18cm}
\caption{
Same as Fig. 9 but for the dispersions of ice abundances due to
dispersions in both  $T_{\rm dust}$ and gas-phase chemical abundance.
}
\label{Figure. 11}
\end{figure*}

\subsection{What causes ice  abundance spreads ?}

\subsubsection{$T_{\rm dust}$}

As shown in Fig. 9,
there is a weak negative radial gradient in the  mean  ${\rm H_2O}$ ice abundances
at $R>3$ kpc (i.e., outside the central
bulge) for the fiducial model. The adopted negative metallicity gradient
in the present disk model is responsible for the more efficient ${\rm H_2O}$ ice
formation in the inner regions with higher metallicities ($Z \approx 2Z_{\odot}$).
Clearly, the dispersions of $T_{\rm dust}$ in radial bins do not introduce
large dispersions in the mean ${\rm H_2O}$ abundances, in particular, at $R>5$ kpc.
This is mainly because ${\rm H_2O}$ ice formation is determined by
oxygen abundances and is much less sensitive to
$T_{\rm dust}$. 
These results imply that there can be smaller differences in ${\rm H_2O}$ ice abundances between
different molecular cores at different locations within a galaxy.
The central dip-like feature in the radial profile
is due largely to the strong radiation field of the central compact bulge and bar,
which can significantly increase $T_{\rm dust}$ at $R<3$ kpc:
this strong UV radiation, however, cannot directly influence molecule formation
owing to severe dust extinction within molecular cores.
It should be noted here that the Galactic central region is observed to be rich in gas- and ice-phase molecules
and organic matters.

Such a negative gradient of ice abundances can be more clearly seen in the radial profile
of mean ${\rm CO_2}$ abundances. The ${\rm CO_2}$ ice abundances
decrease by almost two orders of magnitudes from $R=0$ to $R=5$ kpc,
which means a much steeper radial gradient compared to ${\rm H_2O}$. 
Dispersions of $T_{\rm dust}$ in radial bins  can introduce large dispersion of this ice abundance,
in particular, at $R<5$ kpc, where large $T_{\rm dust}$ variations are found  (see Fig. 5).
This result indicates that a galaxy can show a dramatic variation in ${\rm CO_2}$ ice abundances
between the cores of molecular clouds at different locations within the galaxy. 
The radial variation in the ${\rm CO_2}$-to-${\rm H_2O}$-ratios  will be compared with the corresponding
observations in the Galaxy.
It is possible that even molecular cores with  similar distances from the centre of a galaxy
can have quite different ${\rm CO_2}$ ice abundances.
Like ${\rm H_2O}$ ice,   
the radial profile of ${\rm CO_2}$ ice abundances shows a much less steep gradient and 
smaller dispersion of the abundances in radial bins at $R>10$ kpc due to much less variations
of the physical properties of ISM.

Radial profiles of mean CO and ${\rm CH_3OH}$ ice abundances are similar to that of ${\rm H_2O}$ in the sense
that (i) they have the central dip at $R<3$ kpc and (ii) they have weak negative gradients (i.e.,
smaller in outer regions). Dispersions of these two  ice abundances in radial bins 
due to the dispersions of $T_{\rm dust}$ 
are significantly larger  compared with ${\rm H_2O}$ ice, suggesting that 
the formation of these two ice components is more sensitive to the physical properties of gas and dust that
vary much among different molecular clouds.
These  different levels of dispersions among different four ice abundances 
within a galaxy predicted in the present study 
would need to be observationally investigated and thus confirmed.

\subsubsection{Gas-phase elemental abundances}

Fig. 10 demonstrates that dispersions in gas-phase chemical abundances can cause
a large degree (almost an order of magnitude)  of ${\rm H_2O}$ abundance spreads  in
all radial bins. This result reflects the fact that ${\rm H_2O}$ ice formation
on dust grains is strongly dependent on gas-phase oxygen abundances. Inhomogeneous
mixing of gas-phase oxygen ejected from CCSNe and AGB stars with ISM can end up with
large dispersion in the oxygen abundances of local ISM. Consequently,
${\rm H_2O}$ ice abundances can be quite different between different molecular gas with
different oxygen abundances.
The degrees of ${\rm H_2O}$ abundance spreads in the outer part of the simulated galaxy
do not vary so much with radii due to a very weak radial dependence of oxygen abundance spreads
in ISM.
The ${\rm CO_2}$ ice abundance spreads in radial bins are smaller compared to ${\rm H_2O}$ ice,
which physically means that ${\rm CO_2}$ ice formation is less sensitive to dispersions of
gas-phase oxygen abundances (compared to $T_{\rm dust}$).

Fig. 10 also indicates that the central region of the simulated galaxy ($R<2$ kpc) can have 
a very large dispersion in ${\rm CO}$ ice abundances.  This is mainly because ${\rm CO}$ ice formation
is strongly influenced by carbon
and oxygen abundances, both of which vary to a large degree in the central regions with
more efficient star formation.
Such a large dispersion in the central regions can be seen also in the
radial profile of  ${\rm CH_3OH}$ ice abundances.
These results imply that the central regions of galaxies can have molecular gas with
quite diverse ${\rm CO}$ and ${\rm CH_3OH}$ ice abundances. 
It is intriguing that the radial profile of ${\rm CH_3OH}$ ice has systematically smaller
dispersions at $R>3$ kpc compared to ${\rm CO}$ ice.
It is also confirmed that gas-phase ${\rm H_2O}$, ${\rm CO_2}$, CO, and ${\rm CH_3OH}$ abundances
show flat radial profiles for $R>2$ kpc, which are significantly different from the profiles of their
ice-phase abundances.

The physical explanations for ${\rm H_2O}$ and ${\rm CO_2}$ ice abundances
depending on $T_{\rm dust}$ and metallicities are as follows.
The ${\rm H_2O}$
 ice abundance is more sensitive to elemental abundances rather than the dust temperature, and 
conversely, the CO2 ice abundance is more sensitive to the dust temperature rather than 
elemental abundances. In our models, ${\rm H_2O}$ ice is
 formed by the two-body reactions on grain surfaces, OH + H $\rightarrow$ ${\rm H_2O}$
and OH + ${\rm H_2}$ $\rightarrow$  ${\rm H_2O}$ + H, 
while ${\rm CO_2}$ ice is formed by the two-body reaction on grain 
surfaces, OH + CO ${\rightarrow}$  ${\rm CO_2}$ + H. 
 The precursor of ${\rm H_2O}$ and ${\rm CO_2}$, OH, is formed by the hydrogenation reaction of atomic O on grain surfaces. 
For two-body reactions on grain surfaces to occur, reactants must meet at the grain 
surface through surface diffusion. The surface diffusion of OH is negligible 
at $< 20$ K (Miyazaki et al. 2022). 
As the surface diffusion of H atoms and ${\rm H_2}$
 is efficient even at 10 K, the ${\rm H_2O}$ ice abundance is 
not so dependent on dust temperature and is largely determined by the elemental abundance of oxygen. 
On the other hand, the surface diffusion of CO becomes efficient at $> 15$ K and 
is more efficient at higher temperatures, the ${\rm CO_2}$ abundance increases with increasing temperature. 
At the temperature higher than $\approx 20$ K, the ${\rm H_2O}$ ice abundance becomes 
lower, because the thermal desorption of atomic H and ${\rm H_2}$
becomes non-negligible, which reduces the formation rate of ${\rm H_2O}$ ice.

\subsubsection{A combination of $T_{\rm dust}$ and chemical abundances}

It is clear in Fig. 11 that a combination of dispersions in $T_{\rm dust}$ and gas-phase
chemical abundances can introduce quite large dispersions of ice abundances in radial bins: the dispersions
are even larger than those derived in Figs. 9 and 10.
These dispersions are more likely to be smaller in the outer parts of the simulated galaxy
for all four ice components. The predicted large dispersion in each radial bin means that 
these ice abundances can be quite different in different molecular gas in a galaxy, even if their distances
from the centre of the galaxy are very similar. 
Given the current scarcity of ice observations across diverse environments and 
galactocentric distances within the Galaxy, observationally verifying these predictions 
remains challenging. The upcoming all-sky near-infrared spectroscopic survey and ice mapping 
by SPHEREx (Ashby et al. 2023) may help test these predictions.

\section{Discussion}
\subsection{Molecular diversity from diverse galaxy environments}

Using the results of the AKARI spectroscopic survey for the LMC,
Shimonishi et al. (2010) have discovered that 
the abundance ratios of ${\rm CO_2}$ to ${\rm H_2O}$ ice toward young massive
stellar objects (YSOs) in the LMC ($\approx 0.36$) are systematically higher 
than those of the Galactic massive YSOs ($\approx 0.17$).
Shimonishi et al. (2016) have also analysed 3.2-3.7 spectrum of LMC's seven YSOs obtained
by VLT and thereby compared 
the abundances of ${\rm CO_2}$, ${\rm H_2O}$, and ${\rm CH_3OH}$ ice 
with those of the Galactic counterparts.
One of key results from this observation is that the ${\rm CH_3OH}$ abundances
of the LMC YSOs are significantly lower than those of the Galactic counterparts
possibly due to the higher dust temperatures of ISM in the LMC.
The deficiency or low abundance of ${\rm CH_3OH}$
 is also reported for several LMC hot cores (Shimonishi et al. 2016b, 2020). Possible deficiency of 
${\rm CH_3OH}$  gas abundances in the LMC as well as the other 
low-metallicity star-forming dwarf galaxy IC10 have been also reported by Nishimura et al. (2016a,b).

Although these observational results have valuable information on what roles galaxy environments
play in the formation of interstellar ice,  the physical interpretations of the results  based on
theoretical modelling of ice abundances in various galaxy environments are yet to be done.
As shown in the present study,  key physical parameters of ISM for interstellar ice formation
such as $T_{\rm dust}$, $R_{\rm UV}$, and $f_{\rm d}$ depend strongly on distances from
galactic centers ($R$), and they are quite different even at a given $R$ in a galaxy.
Therefore, the observed diverse ice abundances can be physically interpreted in the context
of diversity in these physical parameter within galaxies.
Furthermore, given that these parameters can depend on global galactic properties
such as stellar surface mass densities (determining $F_{\rm ISRF}$ and $T_{\rm dust}$),
 star formation rates
($R_{\rm UV}$ and $\zeta_{\rm CR}$), and metal and dust abundance ($f_{\rm dust}$),
the observed ice diversity can be correlated with these global galactic properties.
The present study does not provide specific predictions on the possible
correlations between ice abundances and  global physical properties in galaxies
(e.g., ${\rm H_2O}$ ice mass vs galaxy mass),
because it is clearly beyond the scope of this paper to do so.
We will however discuss these possible correlations for
the observed galaxies
such as the LMC and the SMC
in our future works.

\begin{figure}
\psfig{file=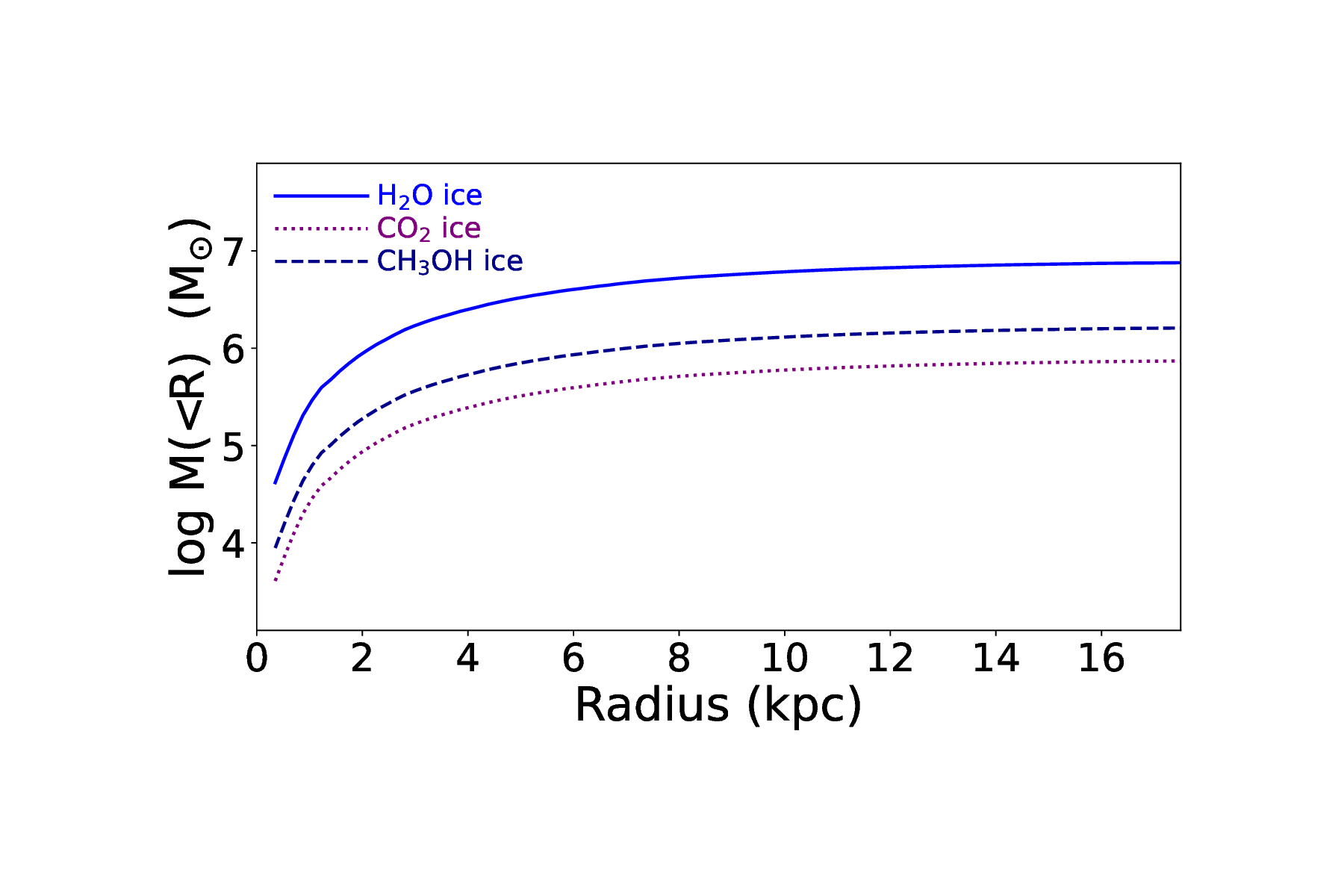,width=8.5cm}
\caption{
Cumulative mass distributions of 
gas-phase CO (blue),  ${\rm H_2O}$ ice (orange), ${\rm CO_2}$ ice (green), 
and ${\rm CH_3OH}$ ice at $T=0.28$ Gyr in the fiducial model.
}
\label{Figure. 12}
\end{figure}

\begin{figure*}
\psfig{file=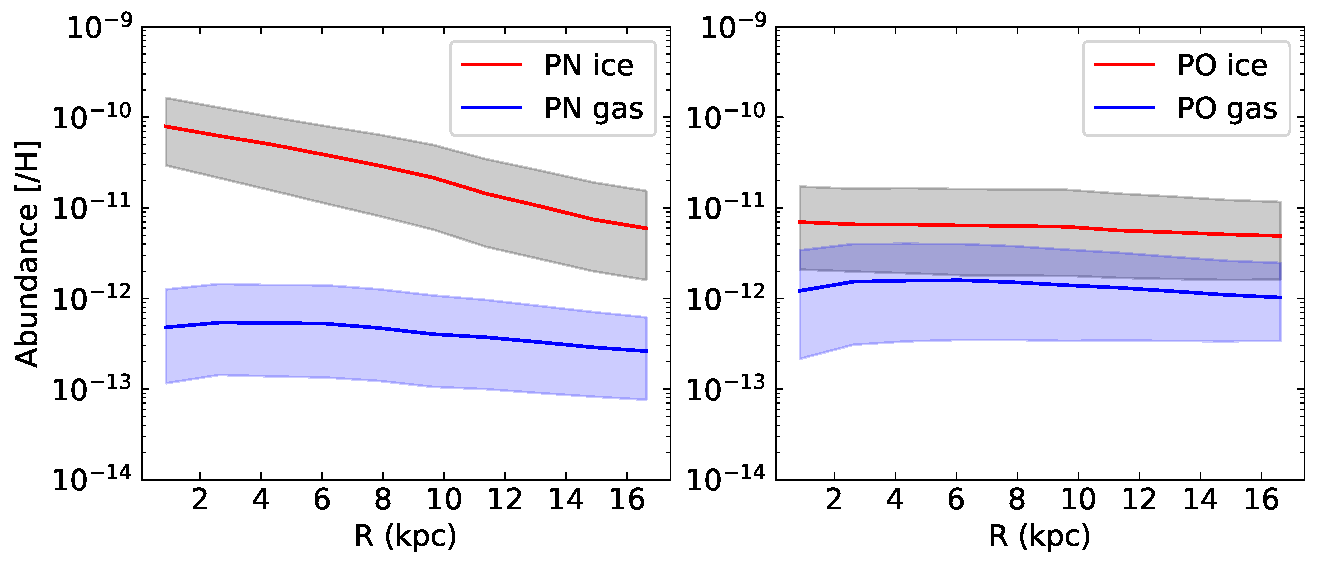,width=18.cm}
\caption{
Same as Fig. 11 but for gas-phase (blue) and ice-phase (red) PN (left) and PO (right).
}
\label{Figure. 13}
\end{figure*}

\subsection{The total ice mass and ice distribution in the Galaxy}

If the  mass of ${\rm H_2O}$ ice  is estimated for {\it all} individual  gas particles
in a simulated galaxy at a given time, then the total mass of ${\rm H_2O}$ ice of the simulated
galaxy ($M_{\rm H_2O}$) can be inferred. 
Since it takes typically several minutes for the adopted astrochemistry code (F15) to derive
the one Myr evolution of interstellar molecules just for a single gas particle,
it is currently impossible to accurately 
estimate ${\rm H_2O}$ abundances for all $N \approx 10^6$ gas particles 
in a simulation within a reasonable timescale (within a few days or so).
Here we roughly estimate $M_{\rm H_2O}$ 
of the Galaxy  using the results predicted from the simulations as follows.
First we calculate the mean or typical  mass fraction of ${\rm H_2O}$ ice ($f_{\rm H_2O}$)
among the selected gas particles
in the fiducial model and then multiply $f_{\rm H_2O}$ with the observed
total ${\rm H_2}$ mass ($M_{\rm H_2}$) of the Galaxy: 
\begin{equation}
M_{\rm H_2O}  = f_{\rm H_2O} M_{\rm H_2},
\end{equation}
where $M_{\rm H_2}$ is observed to be $9 \times 10^8 M_{\odot}$.
It is straightforward to derive the ${\rm H_2O}$ mass fractions of gas particles
from  the number densities  of ${\rm H_2O}$ ice predicted from the present simulations.
Here we estimate  $M_{\rm H_2O}$ and the cumulative mass distribution
by assuming that the number density ratio of ${\rm H_2O}$ ice to H is fixed at  $10^{-4}$ consistent
with our predictions from the present simulation. Similarly,
we estimate $M_{\rm CO_2}$ and $M_{\rm CH_3OH}$ and their cumulative mass distributions.

Fig. 12 shows the cumulative mass distributions of these three ice species
within 17.5 kpc in the fiducial model at $T=0.28$ Gyr. The three profiles all have
steep increases until $R\approx 4$ kpc due to higher fractions of molecular clouds
in the inner regions.
The total masses of ${\rm H_2O}$, ${\rm CO_2}$, and ${\rm CH_3OH}$ ice
within $R=17.5$ kpc (i.e., the stellar disk size)
are  $7.6 \times 10^6 {\rm M}_{\odot}$,
$7.4 \times 10^5 {\rm M}_{\odot}$,
$1.6 \times 10^6 {\rm M}_{\odot}$,
respectively.
It should be stressed here that this estimation is made assuming  fixed ice-to-H-number ratios
in all gas particles for the three ice components, though the number ratios can be quite different in different
molecular clouds. We will make a much more accurate estimation of these ice masses when we have developed
a much faster code for astrochemistry calculations (i.e., new ``emulator'', as discussed later).
The 2D distributions of ${\rm H_2O}$ ice in this model appears to be very  clumpy and similar to
the ${\rm H_2}$ distributions (see Appendix A). \\

\subsection{Distributions of P-bearing molecules}

Recent millimetre observations of the outer ISM of the Galaxy have detected gas-phase 
PN and PO and thereby estimated their abundances (Koelemay et al. 2023). 
A galactocentric distance to the target star-forming region, WB89-621, is estimated to be 
22.6 kpc based on the kinematic distance (Blair et al. 2008), 
while the parallax measurement by the water maser astrometry suggests the galactocentric distance of 
13.21 $\pm$ 1.58 kpc for this region (Hachisuka et al. 2015). 
Since an enough number of CCSNe that can eject P-rich gas cannot be expected far outside the 
solar circle, Koelemay et al. (2023) 
has suggested that neutron-capture processes in AGB stars, which should be ubiquitous 
even in the Galactic outer part, are responsible for the formation of P-bearing molecules in the outer Galaxy.
It should be stressed here that (i) the P yields of AGB stars are not so high as those
from other possible sources like novae
(e.g., Masseron et al. 2020) and (ii) recent chemical evolution models of the Galaxy
have demonstrated that the observed [Fe/H]$-$[P/Fe] relation can be well reproduced only
if P-enrichment by ONe novae is properly included (Bekki \& Tsujimoto 2024).
Therefore, if ONe nova populations dominate P-enrichment in ISM of galaxies, then
ISM chemically polluted by ONe novae from  older binary populations
can contain a significant amount of P-bearing molecules.
Accordingly, P-bearing molecules can be found even in the very outer part of the Galaxy
where star formation is much less active.

Since the present study includes P-enrichment by ONe nova in its chemical evolution models,
it can address the physical origins of possible radial abundance gradients of P-bearing molecules in the Galaxy.
As shown in Fig. 13, the radial gradients of gas-phase PN and PO abundances are pretty flat,
indicating that the PN and PO abundances in the outer regions ($R>15$ kpc)
can be as high as those in the inner regions ($R<5$ kpc). 
Although this lack  of radial gradients would be compared at least qualitatively with
existing (e.g., Koelemay et al. 2023) and future new observations,
the physical origin of this is less clear at this stage.
It is unlikely that P-enrichment by ONe nova for less than 1 Gyr causes
flat abundance gradients of the elements, N, O, and P.
The result  implies   that (i)  gas-phase PN and PO
abundances are not so  sensitive to  initial (negative) radial gradient of N, O, P
and (ii) several competing effects (e.g., $T_{\rm dust}$,  $f_{\rm d}$ etc)
might conspire to keep relatively
constant abundances of these PN and PO molecule.
The physical reason for the flat PO  profile
is that PO can be converted into PN through the following path,
\begin{equation}
{\rm PO + N} \rightarrow {\rm PN+O},
\end{equation}
where both the formation and destruction rates of the P-bearing molecules
depend on chemical abundances of ISM.
Diverse  ISRFs and elemental abundances within molecular clouds cause    
the significant dispersions ($\approx 0.5$ dex) of the gas-phase PN and PO abundances in
each radial bin.
Large spreads of elemental abundances within molecular clouds
are largely responsible for these  dispersions:
Appendix B discusses in detail how dispersions in $T_{\rm dust}$ and
elemental abundances result in PN and PO abundance spreads.

Intriguingly, PN ice abundances are larger 
in the inner regions (i.e., negative radial gradient),
whereas PO ice does not show such a steep radial gradient.
Again, the physical origin of this clear difference in the radial abundance gradients
is unclear at this stage, though PN ice abundances might depend strongly on
the radial gradients of N and P abundances.
Both PN and PO ice show large dispersions in each radial bin, which is due largely
to chemical abundance spreads among molecular clouds within each bin.
These derived radial gradients of P-bearing molecules are currently unfeasible to be compared with
the corresponding observations, simply because the entire Galaxy has not been investigated in terms
of these abundances (Koelmay et al. 2023).
It is thus worthwhile for future observations to map the distributions of PN and PO molecules in ISM
at different radii.  Extensive comparisons between
such observed distributions of P-bearing molecules and corresponding predictions from
our simulations will enable us to
reveal not only how P-bearing molecular can be formed at different radii of the Galaxy
but also whether or not ONe novae are responsible for rich P-bearing molecules in the outer
parts of the Galaxy.

\subsection{Implications on planet formation}

The present study has first demonstrated that the ${\rm CO_2}$-to-${\rm H_2O}$-ice-ratios
can be quite different between different star-forming molecular clouds, which could have 
a profound implication on planet formation and the Galactic habitable zone.
It is theoretically demonstrated that
the ice compositions in the building blocks of planets (i.e., 0.1 $\mu$m-scale dust)
can determine the stickiness of ice particles
 and therefore influence the planet formation from aggregation of dust 
(e.g., Arakawa \& Krijt 2021).
Therefore, if ice produced in star-forming cores of molecular clouds
can be transferred into the proto-planetary disks,
then the present results  imply that planet formation processes could depend on $R$
due to negative radial gradients of  $T_{\rm dust}$ and metallicities  of the Galaxy.
If the formation of rocky planets can be severely suppressed by lower degrees of stickiness of ice-covered
dust particles in GMCs at larger $R$, then the outer Galaxy might contain only a much smaller number of
planets that are habitable.
Thus, ice compositions, which can be quite different at different radii in the Galaxy,
could  be a factor that defines the Galactic habitable zone.

As shown in Fig. 7,  the formation timescale of ${\rm H_2O}$ ice ($t_{\rm H_2O}$) is well
less than $10^6$ yr around the solar neighborhood ($Z \approx Z_{\odot}$).
 If the accretion timescale of oxygen atoms
onto dust grains is a good approximation of  $t_{\rm H_2O}$ (e.g., Hollenbach et al. 2009),
then $t_{\rm H_2}$ of a gas cloud is described as follows:
\begin{equation}
t_{\rm H_2O}  = 8 \times 10^4 (\frac{ f_{\rm o} }{ f_{\rm o, 0} })^{-0.5}
(\frac{ \rho_{\rm g} }{ 10^4 {\rm cm^{-3} } })^{-1}
(\frac{ T_{\rm g} }{ {\rm 30 K} })^{-0.5} {\rm yr},
\end{equation}
where $f_{\rm o}$ is the mass fraction of gas-phase oxygen, 
$f_{\rm o,0}$ is $f_{\rm o}$ of the Sun,
$\rho_{\rm g}$ is the gas density,
and $T_{\rm o}$ is the  gas temperature.
In a typical core of a molecular cloud, $t_{\rm H_2O}$ is much shorter than
the lifetime ($t_{\rm OB} \approx 3 \times 10^6$ yr)
of the most massive OB stars ($m > 50 {\rm M}_{\odot}$) that can explode as CCSNe, i.e.,
\begin{equation}
t_{\rm H_2O}  < t_{\rm OB},
\end{equation}
which ensures that evaporation and destruction of ${\rm H_2O}$ ice by strong UV radiation
from  OB stars is well proceeded by
water formation within star-forming gas clouds and the subsequent transfer of ${\rm H_2O}$ ice
to planetary systems: this $t_{\rm H_2O}<t_{\rm OB}$ might be a requirement for the formation
of planets with water.
However, it should be stressed here that $t_{\rm H_2}$ can be longer than $t_{\rm OB}$ in a star-forming cloud,
if the metallicity ($Z$) of the cloud is less than
$0.01 Z_{\odot}$ and $\rho_{\rm g}$ is as low as $10^3$ cm$^{-3}$.
Therefore, the planetary systems within such a low-Z and low-density cloud are highly unlikely 
to contain water, which is indispensable for lives of all living creatures on the earth.
We thus  suggest that $t_{\rm H_2O}$ is also a factor that defines the Galactic habitable zone.

\section{Concluding remarks}

We have 
investigated the spatial distributions of the abundances of gas- and ice-phase  molecules 
in galaxies through 
applications of  our astrochemistry code (F15)
to the results of our original galaxy-scale hydrodynamical  simulations.
In  our new postprocessing method, the abundances of interstellar molecules
at the locations of individual gas particles 
can be predicted using
gaseous, dust-to-gas-ratios ($f_{\rm d}$), dust temperatures ($T_{\rm dust}$),
UV radiation strength  normalised to the local value ($R_{\rm UV}$), 
and gas- and dust-phase
elemental abundances of various elements (e.g., C, N, O, P, and Cl).
Since the simulations cannot resolve the gas densities ($\rho_{\rm g}$) of molecular cores ($<1$ pc),
we have adopted a reasonable assumption for $\rho_{\rm g}$).
We can therefore predict how the interstellar molecular properties depend on
local physical properties of ISM and stars and how they evolve with time through various global
dynamical processes of galaxies such as bar/spiral arms formation and feedback effects by CCSNe.
We have focused  particularly on the abundances of gas- and ice-phase CO, ${\rm H_2O}$,
${\rm CO_2}$, and ${\rm CH_3OH}$ molecular species  in galaxies that are very similar 
to our Milky Way Galaxy. The preliminary results are as follows:

(1) The simulated star-forming disk galaxy
shows (i) significant radial ($R$) gradients of $T_{\rm dust}$, $f_{\rm d}$,
gas-phase chemical abundances (e.g., $f_{\rm o, g}$ for oxygen), and $R_{\rm UV}$
and (ii) large variations in these physical properties even for a given $R$.
Furthermore, $R_{\rm UV}$, $f_{\rm d}$, and gas-phase chemical abundances
can be quite different among molecular clouds for a given $T_{\rm dust}$.
These diverse physical conditions of galaxy environments can end up with
diverse abundances of interstellar molecules within molecular 
clouds located at different regions of a galaxy. \\

(2) The radial profiles of mean abundances of ${\rm H_2O}$, CO, ${\rm CO_2}$, and ${\rm CH_3OH}$ ice
show negative gradients (i.e., larger in inner regions) at $R>3$ kpc  in the simulated galaxy.
The gradient of ${\rm CO_2}$ ice is  quite steep at $R<5$ kpc, showing by almost two orders of magnitude
differences between different regions with $0<R<5$ kpc. The four ice components
show relatively shallow yet negative gradients at $R>5$ kpc, where star formation and the resultant
chemical enrichment proceed more slowly compared to the inner regions.
As a result of different radial variations of ${\rm H_2O}$ and ${\rm CO_2}$ ice abundances,
the ${\rm CO_2}$-to-${\rm H_2O}$ ice abundance ratios can dramatically vary with radii such that
they can be quite low ($<0.01$) at $R>5$ kpc in the simulated disk galaxy.
Such large variations in  ${\rm CO_2}$-to-${\rm H_2O}$ ice abundance ratios will need to be tested
against observations for various molecular clouds located at different radii in the Galaxy. \\

Large dispersions in  $T_{\rm dust}$ of molecular clouds
are responsible for  a large degree of ${\rm CO_2}$ ice abundance spreads
at different radii in the simulated disk  galaxy.
The degree of  abundance spread is larger
in the inner region of the galaxy, where $T_{\rm dust}$ dispersions are larger.
These results reflect the fact that the formation processes of ${\rm CO_2}$ ice
are very sensitive to $T_{\rm dust}$ of molecular clouds.
The degree of ${\rm H_2O}$ ice abundance spreads caused by $T_{\rm dust}$
dispersions is, on the other hand, overall small,
because the formation of this ice does not depend so strongly on $T_{\rm dust}$.
Radial profiles of CO and ${\rm CH_3OH}$ ice abundances have large spreads 
at $R<5$ kpc, which means that the formation of the two ice components 
is sensitive to $T_{\rm dust}$.
These four ice components show smaller degrees of abundance spreads in the outer parts of
their radial profiles. \\

(4) 
Simulated disk galaxies
show the central dips in the radial profile of mean ${\rm H_2O}$,
${\rm CO}$, and ${\rm CH_3OH}$  ice abundances.
This is due largely  to the 
strong radiation field of the central high-density
 bulge component that can heat up the dust thus
suppress the formation of ${\rm H_2O}$ ice due to higher $T_{\rm dust}$.
This results implies that ${\rm H_2O}$ ice formation in the inner regions of galaxies
can be influenced by the properties of bulges, such as the age and metallicities of 
their stellar populations and the mass fractions of the bulges,
because these properties can determine the strengths of ISRFs in the inner region of galaxies. 
It is our future study to investigate the abundances of these molecules in disk galaxy models
with different bulge-to-disk-ratios and Hubble types.
\\

(5) Dispersions in  gas-phase chemical abundances  of molecular clouds
can also  cause large degrees of ice abundance spreads, in particular,
${\rm H_2O}$ ice.
The abundances of ${\rm H_2O}$ ice can differ by almost an order of magnitude between different
molecular clouds at a similar $R$.
This is mainly because the formation processes of ${\rm H_2O}$ ice on dust grains
are more sensitive to oxygen abundances of ISM for a given $T_{\rm dust}$.
Such large abundance spreads can be also found for CO and ${\rm CH_3OH}$ ice in the inner regions
of the simulated galaxy ($R<3$ kpc).
The radial dependence of these abundance spreads due to dispersions of
gas-phase chemical abundances cannot be so small in the outer parts of the galaxy ($R>10$ kpc),
which is in a striking contrast with the abundance spreads due to $T_{\rm dust}$ dispersions.\\

(6) The combination of large dispersion in
$T_{\rm dust}$ and chemical abundances of molecular clouds
can cause large degrees of  ${\rm H_2O}$, ${\rm CO_2}$, CO and ${\rm CH_3OH}$ ice abundance spreads
for a given $R$.
The ice-phase ${\rm CO_2}$-to-${\rm H_2O}$-ratios can be significantly different
in different $R$ in a galaxy due to the large dispersions of $T_{\rm dust}$ and
chemical abundance patterns at different $R$.
This difference in ice-ratios at different radii ($R$)  from the Galactic centre is suggested
to possibly influence the details of planet formation from aggregation of ice-covered dust particles
within different star-forming molecular clouds at different $R$.
\\

(7) The total masses of ${\rm H_2O}$, ${\rm CO_2}$, and ${\rm CH_3OH}$ ice
($M_{\rm H_2O}$, $M_{\rm CO_2}$, and ${\rm CH_3OH}$, respectively) in galaxies can be predicted
from number ratios of these ice components to hydrogen atom in molecular clouds
derived from the present simulations.
For (fixed) typical ice-to-H-ratios 
$M_{\rm H_2O}$, $M_{\rm CO_2}$, and ${\rm CH_3OH}$ in the Galaxy are
predicted to be $7.6 \times 10^6 {\rm M}_{\odot}$,
$7.4 \times 10^5 {\rm M}_{\odot}$,
$1.6 \times 10^6 {\rm M}_{\odot}$, 
respectively. The 2D distributions of these ice in galaxies are clumpy and similar to
the ${\rm H_2}$ distributions. \\

(8) The dust size is fixed at a typical  value (i.e., 0.1$\mu$m) for all gas particles at different
locations 
in the present simulations,
though ice formation processes and $T_{\rm dust}$ evolution  can be quite
 different between different dust sizes.
Also the adopted assumption of $\zeta_{\rm CR}$ depending only galaxy-scale star formation rates
would be less realistic in the sense that local variation of $\zeta_{\rm CR}$ is completely ignored 
in the present study.
Thus we need to overcome these modelling problems
to predict the abundances of interstellar molecules
in our future works: more discussion on this is given
in Appendix C\\

\section{DATA AVAILABILITY}
The data used in this paper (outputs from computer simulations) 
will be shared on reasonable request
to the corresponding author.

\section{Acknowledgment}
We are  grateful to the referee  for  constructive and
useful comments that improved this paper.
T.S. and K.F. acknowledge support from JSPS KAKENHI (Grant Numbers JP20H05845 and JP21H01145).
The test simulations of the adopted new code
were  performed on the OzSTAR national facility at Swinburne University of Technology. 
The OzSTAR program receives funding in part from the Astronomy National 
Collaborative Research Infrastructure Strategy (NCRIS) allocation provided by the 
Australian Government, and from the Victorian Higher Education State Investment 
Fund (VHESIF) provided by the Victorian Government.

{}

\appendix

\section{2D maps of physical properties}

Figs. A1 and A2 show that the 2D distributions of dust surface densities
($\Sigma_{\rm dust}$) and gas-phase oxygen abundances ($\Sigma_{\rm o, g}$) at a given time step
in the fiducial model
are very similar with each other in the sense that both show higher surface densities 
in the spiral arms and the central bar. Spiral arms and bars can compress ISM to form
high-density gaseous regions, where dust grains can grow due to accretion of gas-phase metals
onto dust. Consequently, $\Sigma_{\rm dust}$ can be significantly higher within spiral
arms compared with
the inter-arm regions in the disk galaxy.
Although accretion of oxygen atoms onto dust grains can reduce the gas-phase oxygen abundances,
such reduction is not so dramatic as to alter the global kpc-scale distribution of gas-phase
oxygen abundances.
Clearly the time evolution of 2D $\Sigma_{\rm dust}$ and $\Sigma_{\rm o, g}$ distributions just follows
that of 2D $\Sigma_{\rm g}$, which confirms that  dynamical evolution of the galactic disk
can determine the global distributions of gas, dust, and metals.

It is infeasible for the present study to derive ${\rm H_2O}$ ice abundances from  all 
1 million gas particles in a simulation at a given time step due to the required large amount
of time for astrochemical calculations of more than 400 molecules.
Therefore, the present study cannot self-consistently predict the 2D distributions of ${\rm H_2O}$
ice abundances in galaxies.
However, if 
we assume a fixed ${\rm H_2O}-$to$-{\rm H}$ abundance ratio ($R_{\rm H_2O}$),
we can derive a {\it possible}  2D distribution of water ice ($\Sigma_{\rm H_2O}$ in the fiducial model.
Fig. A3  describes the time evolution of the $\Sigma_{\rm H_2O}$ distribution in
the model with $R_{\rm H_2O}=10^{-4}$: it should be noted here that we assumed this fixed
value just for the purpose of illustrating the {\it possible}  $\Sigma_{\rm H_2O}$ distribution 
in the disk galaxy, though we know that such a fixed value is unrealistic.
The simulated $\Sigma_{\rm H_2O}$ distributions show clumpy
structures as well as global spiral-arm patterns, and also appear to follow more closely 
the $\Sigma_{\rm UV}$ (or ${\rm H_2}$) distribution, which indicates that $\Sigma_{\rm H_2O}$
distributions can be similar to those of star-forming regions in galaxies.
As discuss in the main text,
our future simulations with a fully self-consistent model for ice feedback effects will be
able to predict more accurate galaxy-scale distributions of ${\rm H_2O}$ ice as well as those
of other major ice species such as ${\rm CO_2}$ and ${\rm CH_3OH}$.

\begin{figure*}
\psfig{file=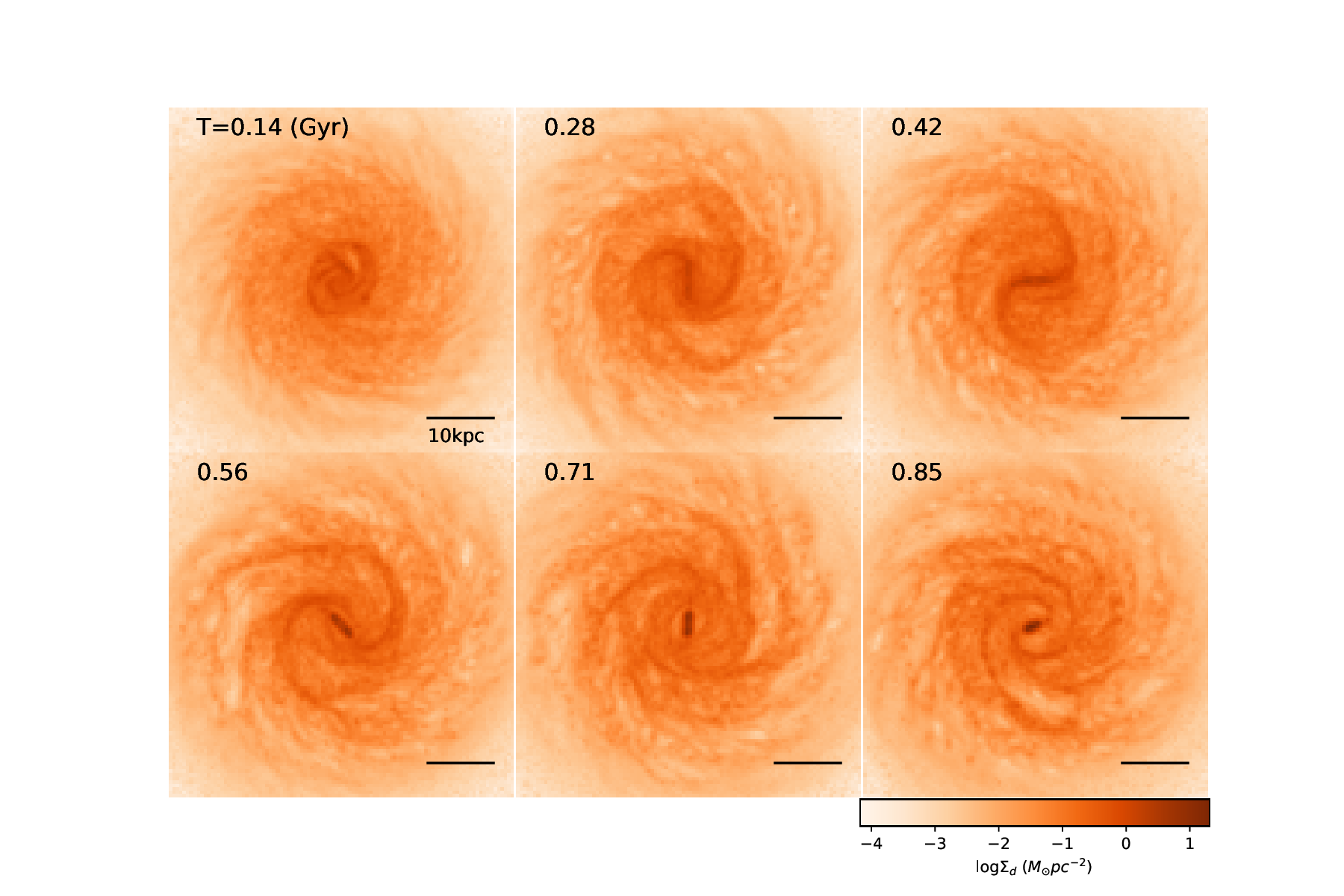,width=19.0cm}
\caption{
Same as Fig. 2 but for the surface densities of dust  ($\Sigma_{\rm dust}$).
}
\label{Figure. 14}
\end{figure*}

\begin{figure*}
\psfig{file=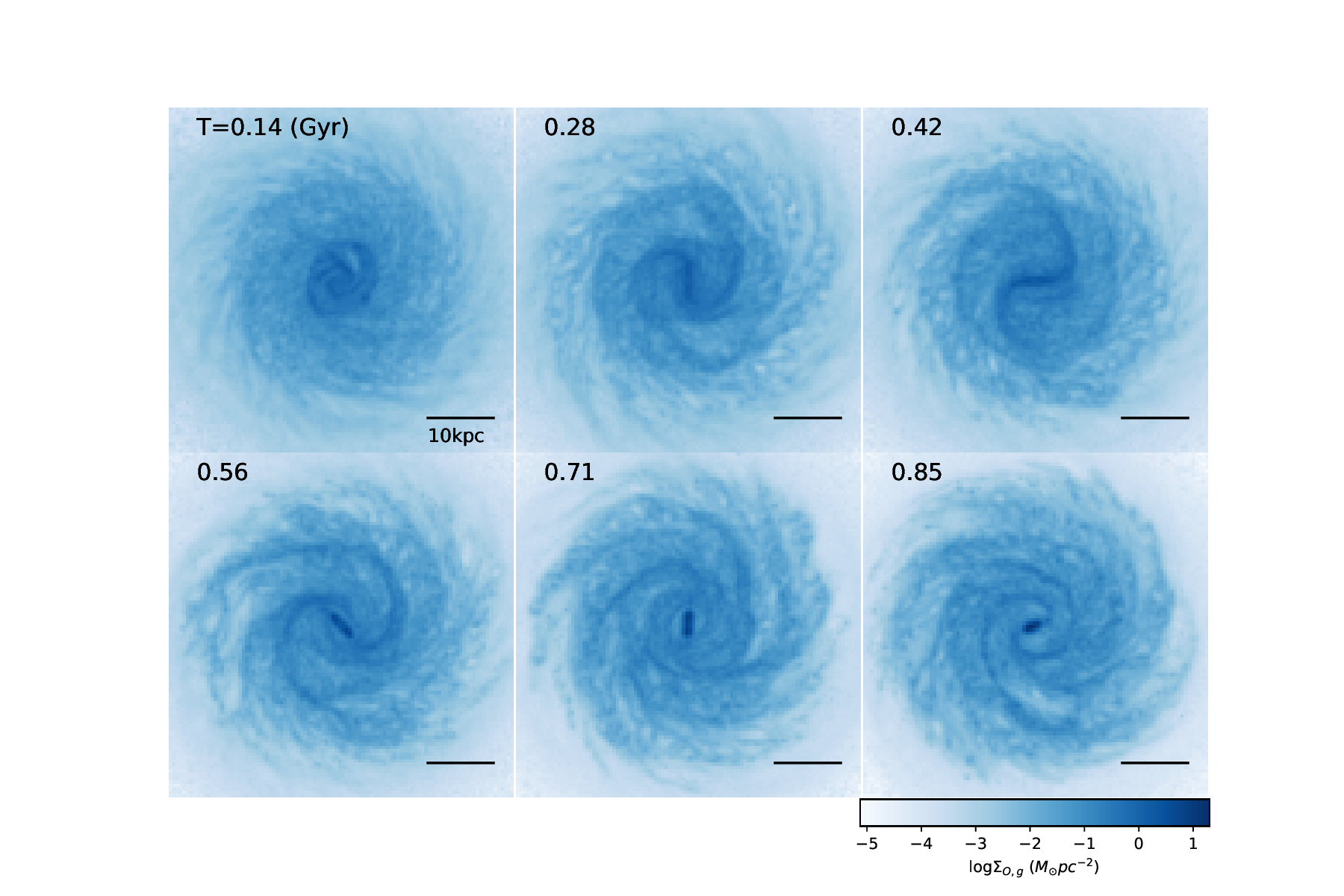,width=19.0cm}
\caption{
Same as Fig. 2 but for the  surface densities  of gas-phase oxygen  ($\Sigma_{\rm o,g}$).
}
\label{Figure. 15}
\end{figure*}

\begin{figure*}
\psfig{file=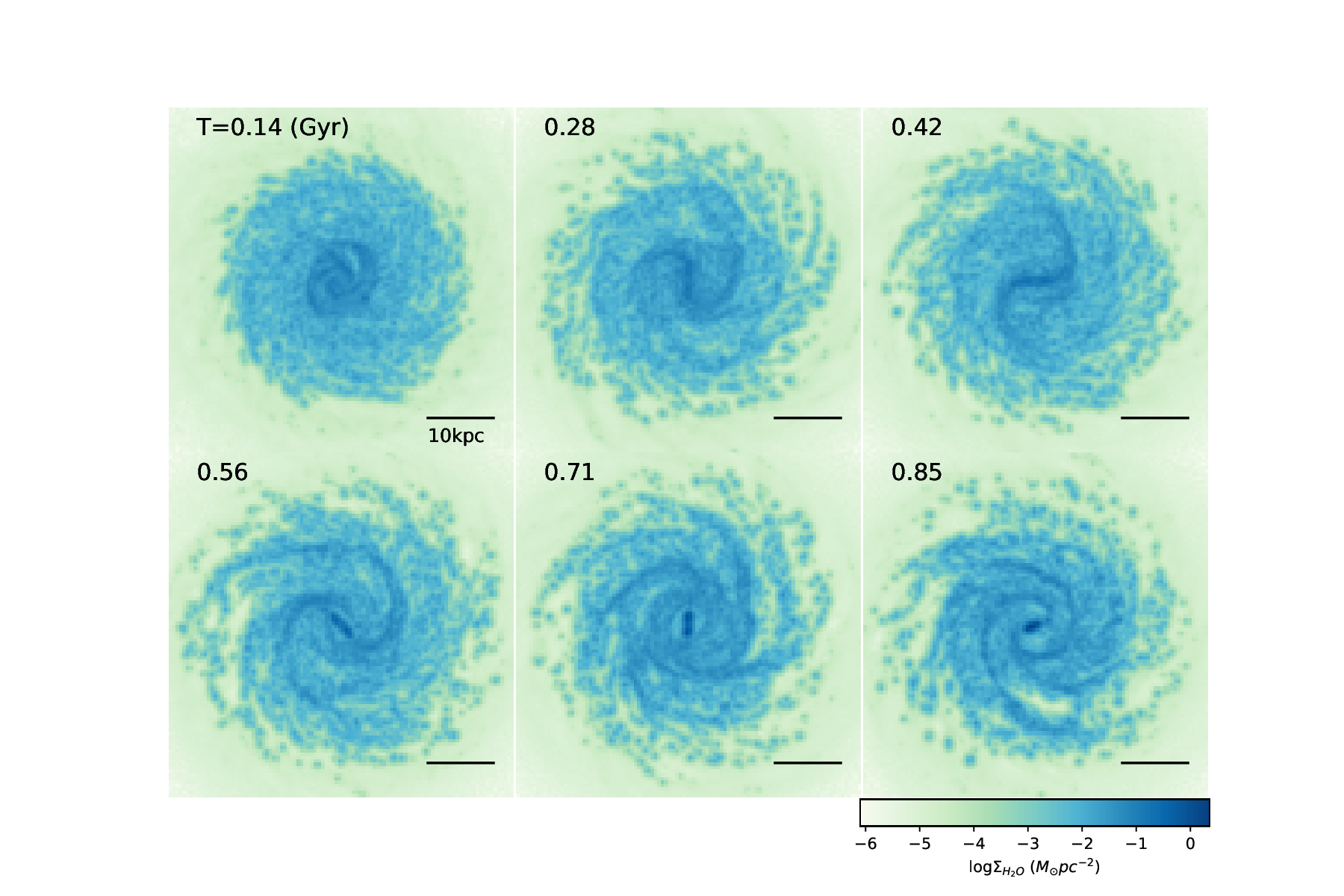,width=19.0cm}
\caption{
Same as Fig. 2 but for the  surface mass densities  of ${\rm H_2O}$ ice  ($\Sigma_{\rm H_2O}$).
The abundance ratio of ${\rm H_2O}$ ice to H is fixed at $10^{-4}$ in all molecular
gas so that the 2D map can be derived from the simulated galaxy: the 2D map is not based
on astrochemical calculations of ${\rm H_2O}$ abundances for all individual
molecular clouds using the physical
properties of the clouds (e.g., $T_{\rm dust}$).
We are now developing to develop a ``emulator'' which can calculate the abundances of these
ice components are much faster than the fully astrochemistry code adopted in the
present study. Accordingly, our future simulations with such an emulator will
be able to derive much more realistic 2D maps of ${\rm H_2O}$ ice.
}
\label{Figure. 16}
\end{figure*}

\section{Radial profiles of PN and PO abundances dependent on $T_{\rm dust}$ and 
metallicities}

\begin{figure*}
\psfig{file=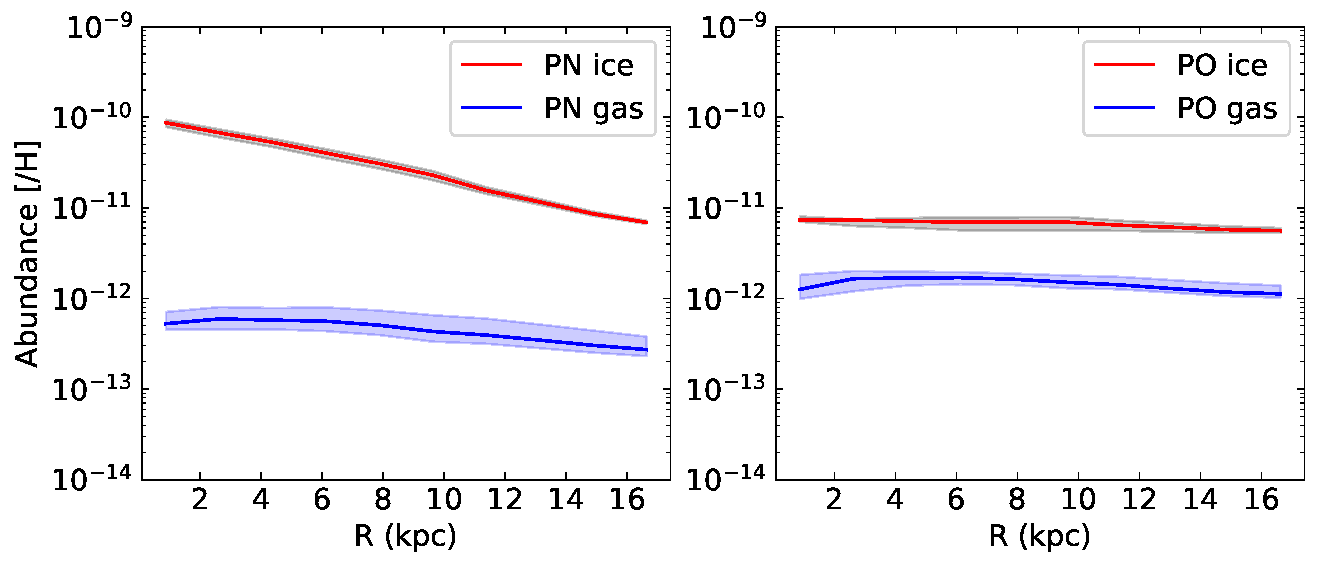,width=19.0cm}
\caption{
Same as Fig. 9 but for PN and PO abundances.
Gas-phase (blue) and ice-phase  (red)  abundances are shown in a same frame so that 
differences in radial profiles between the two can be  readily identified.
}
\label{Figure. 17}
\end{figure*}

\begin{figure*}
\psfig{file=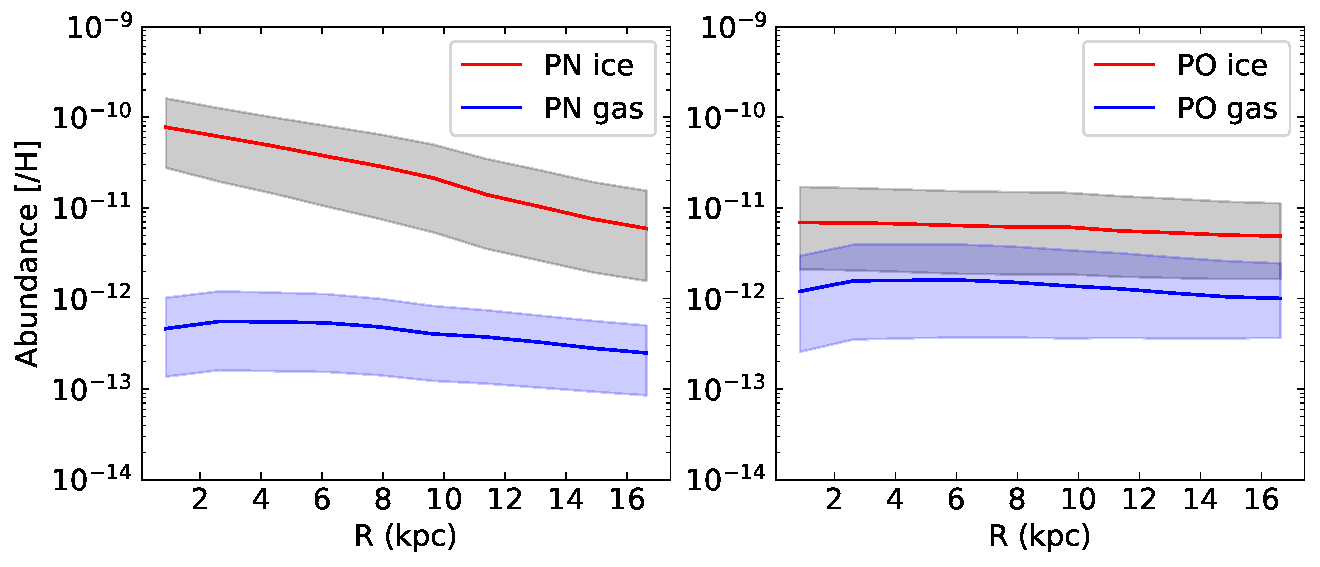,width=19.0cm}
\caption{
Same as Fig. 10 but for PN and PO abundances.
}
\label{Figure. 18}
\end{figure*}

Fig. B1 describes how $T_{\rm dust}$ dispersions in different molecular clouds
influence the radial profiles of gas-phase and ice-phase PN and PO abundances 
in the fiducial disk galaxy model. Clearly the abundances of these  P-bearing molecules 
are much less sensitive to $T_{\rm dust}$ at all radii in this model, which indicates
that diverse strengths of ISRFs in ISM, which can determine $T_{\rm dust}$,
are not a key determinant of these abundances. As discussed in the main text,
the radial profiles of gas-phase PN abundances are very flat due to the weak dependence of 
the formation processes of gas-phase PN 
on radial variations of ISM properties.
The ice-phase PN, on the other hand, clearly shows a negative radial gradient in the sense
that the inner regions with higher P and N abundances (and stronger $F_{\rm UV}$)
can have higher abundances of ice-phase PN. 
Although the PN ice abundances are by almost two orders of magnitudes higher
than their gas-phase abundances at $R<5$ kpc,  such gas- and ice-phase
abundance differences become  much less remarkable
in the outer parts ($R>10$ kpc)  of the disk.
The derived  rather small spreads in PN abundances due to  $T_{\rm dust}$ dispersions
can be clearly seen in gas-phase and ice-phase PO abundances,

Fig. B2 demonstrates that elemental abundance spreads can cause large
dispersions in both gas-phase and ice-phase PN and PO abundances for each radial bin.
These dispersions do not depend on radii, and there is no/little difference in
these dispersions between gas-phase (and ice-phase) PN and PO abundances.
The magnitudes of these dispersion in molecular abundances 
are  very similar to those
shown in Fig. 13 for a model that incorporates spreads in $T_{\rm dust}$ and elemental abundances.
Therefore, these results in B1, B2, and Fig 13 demonstrate that
spreads in elemental abundances are the major reason for the large dispersions in PN and PO
molecular abundances shown in Fig. 13.
The weak dependence of PN and PO abundance dispersions on $T_{\rm dust}$ would be surprising,
however, the simulated spreads in $T_{\rm dust}$ among molecular clouds
might not be large enough to cause significant dispersion in the abundances.
The present study has investigated only one disk galaxy model in which
the initial radial gradients of elemental abundances are fixed. 
Such radial gradients can be significantly more diverse 
in real disk galaxies
with different luminosities, bulge-to-disk ratios,
gas mass factions, and  morphological types. Thus we will need to investigate
how the radial gradients of PN and PO abundances depend on the radial gradients of
gas-phase elemental abundances in these disk galaxies in order to have 
a more comprehensive picture on the origins of these molecular abundances.

\section{Further improvement of our models}

Although we have first derived the spatial distributions of gas- and ice-phase  molecules in galaxies by
incorporating interstellar chemistry in galaxy-scale hydrodynamical simulations,
the adopted models to derive the physical parameters of ISM (i.e., initial inputs
for interstellar chemistry) such as $T_{\rm dust}$ and $R_{\rm UV}$
are less realistic and somewhat idealised. We therefore need to improve these models
on dust properties, local cosmic ray strengths, and  ice formation on dust grains
in order to more accurately predict the details of the molecular distributions within galaxies. 

\subsubsection{Dust size distributions}
Dust sizes can evolve with time in galaxies through dust growth, destruction, and shattering 
and their evolution processes depend strongly on galactic star formation rates
(e.g., Asano et al. 2013; Nozawa et al. 2015).
Also, dust sizes can influence the abundances of CO and ${\rm H_2}$ gas in  ISM of galaxies
(e.g. Hirashita \& Harada 2017; Hirashita 2023), because
they can determine the reaction rates of these molecules and the shielding effects of dust that
are important for molecular gas formation.
In spite of these importance of dust sizes, we have calculated the abundances of interstellar
molecules by assuming a fixed dust size ($0.1\mu$m).
Since ice-phase chemistry depends also on dust sizes (F15), 
we first need to estimate molecular abundances for different dust grain sizes
and then sum up all contributions from these dust grains.
Given that recent hydrodynamical simulations have already incorporated incorporate dust size distributions
(e.g., Romano et al. 2022), 
it would not be a difficult task for our future study to implement dust size evolution
and thereby discuss how the evolution can influence interstellar chemistry within molecular clouds.

\subsubsection{Dust temperature ($T_{\rm dust}$)}

Both the abundances of ${\rm H_2O}$, ${\rm CO_2}$, and ${\rm CH_3OH}$ ice species
and these abundance ratios  in molecular clouds
depend strongly on $T_{\rm dust}$ (F15), which implies that
these abundance information can be used to infer $T_{\rm dust}$.
In the present study, $T_{\rm dust}$ at each gas particle is estimated from
local ISRF that is derived from age and metallicity distributions 
of stars around the particle and dust extinction.
Although the carbon and silicate dust fractions can change with time due to chemical enrichment
by CCSNe, SNIa, and AGB stars,  they were assumed to be fixed at reasonable values
to predict $T_{\rm dust}$ from ISRF. 
Since the present simulations investigate only $\approx 0.8$ Gyr evolution of galaxies, such an
assumption of fixed dust compositions would not lead to gross under- or over-estimation of 
$T_{\rm dust}$ from ISFR in the present study.
However, $T_{\rm dust}$ needs to be derived from ISRF and time-evolving  carbon and silicate dust abundances
in cosmological
simulations of galaxies in which  the dust abundances change with time significantly.

Time evolution of $T_{\rm dust}$ is quite different between dust grains with different sizes for
a given $F_{\rm ISRF}$ (e.g., Draine 2009). 
Therefore, if dust size distributions are incorporated in our future simulations of galaxies,
then $T_{\rm dust}$ needs to be estimated for individual dust grains with different sizes.
If such $T_{\rm dust}$ differences between different dust grains
are properly considered in astrochemistry calculations for a molecular cloud in a galaxy,
the total mass of ice within the cloud  can be estimated more accurately.
However, it should be very time consuming to estimate the total ice mass in a molecular cloud,
because total ice masses need to be estimated for all individual dust grains within the cloud to integrate
the masses from all grains.
To avoid such an unfeasible amount of time required for astrochemistry calculations,
a small number ($<10$) of representative dust sizes (e.g., $0.01 \mu$m, 0.03$\mu$m, 0.1$\mu$m etc)
and the (discrete) size distribution
function  would need to be introduced in our future simulations of galaxies.
The time evolution of the dust size distributions should be also considered for each individual
molecular clouds in these simulations too, though it is still numerically costly to estimate
the total ice mass in a galaxy.

\subsubsection{Local cosmic ray strength}

The mean star formation rate over  $10^7$ yr for the entire disk of a galaxy
is assumed to determine $\zeta_{\rm CR}$ for all molecular clouds of the galaxy in the present study.
This could be a good approximation for $\zeta_{\rm CR}$, because the star formation rate is proportional
to the rate of CCSNe from which cosmic ray originates. However, $\zeta_{\rm CR}$ can be locally quite strong
in molecular clouds close to the explosions of CCSNe,
whereas $\zeta_{\rm CR}$ can be rather weak where star formation is much  less active.
Such local variations of $\zeta_{\rm CR}$ can possibly increase the degree of diversity in molecular
abundances within galaxies, because gas-dust chemistry depends strongly on $\zeta_{\rm CR}$.
Given that locally different $\zeta_{\rm CR}$ has been already incorporated in recent galaxy-scale
simulations,
it would be straightforward to implement the effects of locally
different $\zeta_{\rm CR}$ on interstellar chemistry in our future simulations.

\subsubsection{Ice feedback effects}

Accretion of interstellar metals (e.g., C atoms) onto the surfaces of dust grains
within cold molecular clouds
has been considered to be  the major mechanism of dust growth in galaxies and 
thus modelled to explain the dust abundances of galaxies (e.g., Dwek 1998: Asano et al.).
The increase of surface areas of dust grains due to this dust growth could furthermore
enhance the formation efficiency  of ice on dust grains per unit time.
However, Ferrara et al. (2016) showed that
ice mantles formed around the refractory cores of dust grains
can prevent interstellar metals from getting contact with 
the surface of the cores to finally suppress the dust growth by metal accretion.
They therefore suggested that dust growth in cold molecular clouds assumed in previous
theoretical models of dust evolution in galaxies is problematic.
Severe suppression of silicate dust growth in  molecular clouds at high redshifts ($z$)
has been also demonstrated by Ceccarelli et al. (2018) in which the thickness of icy mantles
for various molecular species are investigated with the proper physical conditions of ISM in
the high-$z$ universe.

This possibly negative feedback effect of ice formation on dust growth 
(referred to as ``ice feedback'') needs to be 
implemented in galaxy-scale simulations in which dust mass evolution in galaxies
is investigated in detail.
Although the present simulations can predict the total ice mass formed on each dust grain
at a given time step,
such predictions is  made in the post-processing of the simulation data, i.e., only after
the completion of simulations. We adopt this post-processing method, mainly because
the required time 
($\Delta T_{\rm chemi}$) to  calculate the abundances of various interstellar molecules for
all particles at all time steps is expected to be extremely long.  
The typical $\Delta T_{\rm chemi}$ for a simulation is estimated as follows:
\begin{equation}
\Delta T_{\rm chemi} = \frac{ \Delta t_{\rm chemi}  T_{\rm end} N_{\rm gas} }{ \Delta t_{\rm max} },
\end{equation}
where $\Delta t_{\rm chemi}$, $T_{\rm end}$, $N_{\rm gas}$, and $\Delta t_{\rm max}$ are
the time required for astrochemistry calculation just for one gas particle,
the final time (where $T=0$ corresponds to the starting time  of the simulation),  
the total number of gas particles,
and the maximum timestep width.
Accordingly, if 
$\Delta t_{\rm chemi} =100$ seconds, 
$T_{\rm end}=1$ Gyr, 
$N_{\rm gas}=10^6$, 
and $\Delta t_{\rm max}=10^6$ yr are adopted,
then $\Delta T_{\rm chemi}=10^{11}$ seconds ($\approx 3200$ years),
which is far longer than the required time for other astrophysical calculations.

Clearly, it is currently impossible and unrealistic to perform such a long simulation
($\Delta T_{\rm chemi} \approx 10^{11}$ seconds).
This  $\Delta T_{\rm chemi}$ problem could be alleviated if $\Delta t_{\rm chemi}$ is
dramatically shortened (like $10^{-4}$) and if astrochemistry calculations
are done every $10^7-10^8$ years instead of $10^6$ years in simulations. 
A way to reduce $\Delta t_{\rm chemi}$ dramatically is to develop an ``emulator'' in which
the abundances of various molecules can be predicted directly from the input parameters
without detailed chemical calculations (Furuya et al. 2024, in preparation).
Although the predicted abundances are just the approximations of those from full chemical calculations,
they can be used to discuss global galaxy-scale distributions of various molecules.
The new ice feedback 
can determine  dust mass evolution in ISM and thus the effectiveness of dust-related 
feedback effects on ISM (e.g., photo-electric heating and dust-cooling).
Furthermore this new ice feedback has been completely ignored in previous simulations of galaxy
formation and evolution too.
Thus we will use the new emulator for astrochemistry calculations and thereby
implement ice feedback effects in our future simulations to discuss
how mutual interplay between dust and ice can determine their abundances in galaxies.

\end{document}